\documentclass[longauth]{aa}
\usepackage{graphicx,txfonts}
\usepackage{hyperref}

\newcommand\ions[2]{#1$\;${\textsc{#2}}}

\newcommand{\hb}{H$\beta$}
\newcommand{\ha}{H$\alpha$}
\newcommand{\NDWFS}{NDWFS J1425+3254}
\newcommand{\kms}{\hbox{\rm km\,s$^{-1}$}}
\newcommand{\0}{\phantom{0}}

\newcommand{\Oiii}{\hbox{[\ions{O}{iii}]}}
\newcommand{\oiii}{[\ions{O}{iii}]\,$\lambda\lambda4959,\linebreak[0]5007$}
\newcommand{\oiiib}{\hbox{[\ions{O}{iii}]}\,\ensuremath{\lambda4959}}
\newcommand{\oiiia}{\hbox{[\ions{O}{iii}]}\,\ensuremath{\lambda5007}}
\newcommand{\nii}{\hbox{[\ions{N}{ii}]}\,\ensuremath{\lambda\lambda6548,\linebreak[0]6584}}
\newcommand{\niia}{\hbox{[\ions{N}{ii}]}\,\ensuremath{\lambda6584}}
\newcommand{\niib}{\hbox{[\ions{N}{ii}]}\,\ensuremath{\lambda6548}}
\newcommand{\Nii}{\hbox{[\ions{N}{ii}]}}

\newcommand{\Msol}{\hbox{M$_\odot$}}
\newcommand\micron{\mbox{$\mu$m}}

\begin{document}

\title{JWST's PEARLS: A $z\simeq6$ quasar in a train-wreck\\galaxy merger system}
\titlerunning{A $z\simeq6$ quasar in a train-wreck galaxy merger system}
\authorrunning{M.~A.\ Marshall et al.}

\author{Madeline A.\ Marshall\thanks{Email: \href{mailto:mmarshall@lanl.gov}{mmarshall@lanl.gov}}\inst{1} \and Rogier A.\ Windhorst\inst{2} \and Giovanni Ferrami\inst{3,4} \and S.\ P.\ Willner\inst{5} \and Maria del Carmen Polletta\inst{6,7} \and William C. Keel\inst{8} \and Giovanni G. Fazio\inst{9} \and Seth H.\ Cohen\inst{2} \and Timothy Carleton\inst{2} \and Rolf A.\ Jansen\inst{2} \and Rachel Honor\inst{2} \and \\ 
Rafael {Ortiz~III}\inst{2} \and
Jake Summers\inst{2} \and Jordan C.\ J.\ D'Silva\inst{10,4} \and Anton M.\ Koekemoer\inst{11} \and Dan Coe\inst{11,12,13} \and  \\
Christopher J.\ Conselice\inst{14} \and
Jose M. Diego\inst{15} \and Simon P.\ Driver\inst{10} \and Brenda Frye\inst{16} \and Norman A.\ Grogin\inst{11} \and Nor Pirzkal\inst{11}  \and \\
Aaron Robotham\inst{10} \and
Russell E.\ Ryan, Jr.\inst{11} \and Christopher N.\ A.\ Willmer\inst{17} \and Haojing Yan\inst{18} \and Massimo Ricotti\inst{19} \and \\
Adi Zitrin\inst{20} \and 
Nathan J. Adams\inst{14} \and Cheng Cheng\inst{21} \and J. Stuart B. Wyithe\inst{22,4} \and Jeremy Lim\inst{23} \and Michele Perna\inst{24}\and \\ 
Hannah \"Ubler\inst{25} \and 
Chris J. Willott\inst{26} \and Gareth Jones\inst{27,28} \and Jan Scholtz\inst{27,28} \and Mira Mechtley\inst{2} }

\institute{Los Alamos National Laboratory, Los Alamos, NM 87545, USA
\and School of Earth and Space Exploration, Arizona State University,
Tempe, AZ 85287-1404, USA
\and School of Physics, University of Melbourne, Parkville, VIC
3010, Australia
\and ARC Centre of Excellence for All Sky Astrophysics in 3
Dimensions (ASTRO 3D), Australia
\and Center for Astrophysics \textbar\ Harvard \& Smithsonian, 
60 Garden Street, Cambridge, MA, 02138, USA
\and INAF, Istituto di Astrofisica Spaziale e Fisica Cosmica Milano,  Via A. Corti 12, I-20133 Milano, Italy
\and Department of Astronomy \& Astrophysics, University of California at San Diego, 9500 Gilman Drive, La Jolla, CA 92093, USA
\and Dept. of Physics and Astronomy, University of Alabama, Box 870324, Tuscaloosa, AL 35404, USA
\and Center for Astrophysics | Harvard \& Smithsonian, 60 Garden St., Cambridge, MA 02138 USA
\and International Centre for Radio Astronomy Research (ICRAR) and the
International Space Centre (ISC), The University of Western Australia, M468,
35 Stirling Highway, Crawley, WA 6009, Australia
\and Space Telescope Science Institute, 3700 San Martin Drive, Baltimore, MD 21218, USA
\and Association of Universities for Research in Astronomy (AURA) for the European Space Agency (ESA), STScI, Baltimore, MD 21218, USA
\and Center for Astrophysical Sciences, Department of Physics and Astronomy, The Johns Hopkins University, 3400 N Charles St. Baltimore, MD 21218, USA
\and Jodrell Bank Centre for Astrophysics, Alan Turing Building,
University of Manchester, Oxford Road, Manchester M13 9PL, UK
\and Instituto de F\'isica de Cantabria (CSIC-UC). Avenida. Los Castros
s/n. 39005 Santander, Spain
\and Department of Astronomy/Steward Observatory, University of Arizona, 933 N Cherry Ave,
Tucson, AZ, 85721-0009, USA
\and Steward Observatory, University of Arizona,
933 N Cherry Ave, Tucson, AZ, 85721-0009, USA
\and Department of Physics and Astronomy, University of Missouri,
Columbia, MO 65211, USA
\and Department of Astronomy, University of Maryland, College Park, 20742, USA
\and Physics Department, Ben-Gurion University of the Negev, P.O. Box 653,
Be'er-Sheva 8410501, Israel
\and Chinese Academy of Sciences South America Center for Astronomy, National Astronomical Observatories, CAS, Beijing 100101, China
\and Research School of Astronomy and Astrophysics, Australian National University, Canberra, ACT 2611, Australia
\and Department of Physics, The University of Hong Kong, Pokfulam Road,
Hong Kong
\and Centro de Astrobiolog\'{\i}a (CAB), CSIC-INTA, Ctra. de Ajalvir km 4, Torrej\'on de Ardoz, E-28850, Madrid, Spain
\and Max-Planck-Institut für extraterrestrische Physik, Gießenbachstraße 1, 85748 Garching, Germany
\and National Research Council of Canada, Herzberg Astronomy \& Astrophysics Research Centre, 5071 West Saanich Road, Victoria, BC V9E 2E7, Canada
\and Kavli Institute for Cosmology, University of Cambridge, Madingley Road, Cambridge, CB3 0HA, UK
\and Cavendish Laboratory - Astrophysics Group, University of Cambridge, 19 JJ Thomson Avenue, Cambridge, CB3 0HE, UK
}

\abstract
{
We present JWST NIRSpec integral field spectroscopy observations of the $z=5.89$ quasar \NDWFS\ from 0.6--5.3\,\micron, covering the rest-frame ultraviolet and optical at a spectral resolution of $R\sim100$. 
The quasar has a black hole mass of $M_{\rm{BH}}=(1.4\protect\substack{+3.1\\-1.0})\times10^9$\,\Msol\ and an Eddington ratio of $L_{\rm{Bol}}/L_{\rm{Edd}}=0.3\protect\substack{+0.6\\-0.2}$, as implied from the broad Balmer \ha\ and \hb\ lines.
The quasar host has significant ongoing obscured star formation, as well as a quasar-driven outflow with velocity $6050\protect\substack{+460\\-630}$\,km\,s$^{-1}$ and ionised outflow rate of $1650\protect\substack{+130\\-1230}$\,\Msol\,yr$^{-1}$. This is possibly one of the most extreme outflows in the early Universe.
The data also reveal
that two companion galaxies are merging with the quasar host. 
The north-eastern companion galaxy is relatively old and very massive, with a luminosity-weighted stellar age of $65\protect\substack{+9\\-4}$\,Myr, stellar mass of $(3.6\protect\substack{+0.6\\-0.3})\times10^{11}\,M_\odot$, and star-formation rate (SFR) of $\sim$15--30\,$M_\odot$\,yr$^{-1}$.
A bridge of gas connects this companion galaxy and the host, confirming their ongoing interaction.
A second merger is occurring between the quasar host and a much younger companion galaxy to the south, with a stellar age of $6.7\pm1.8$\,Myr, stellar mass  of $(1.9\pm0.4)\times10^{10}$\,\Msol, and SFR of $\sim$40--65$\,M_\odot$\,yr$^{-1}$. 
There is also another galaxy in the field, likely in the foreground at $z=1.135$, which could be gravitationally lensing the quasar with a magnification of $1<\mu<2$ and, thus, $<0.75$\,mag.
Overall, the system is a `train-wreck' merger of three galaxies, with star formation and extreme quasar activity that were likely triggered by these ongoing interactions.
}

\keywords{quasars: supermassive black holes -- quasars: emission lines -- Galaxies: high-redshift -- Galaxies: interactions -- Galaxies: active}

\maketitle

\section{Introduction} 
\label{sec:intro}
Hundreds of quasars within the first billion years of the Universe's history have been discovered using large sky surveys \citep[e.g.][]{fan_2000,fan_2001,Fan2003,willott_2009,Willott2010,Kashikawa2015,Banados2016,Banados2017,Banados2022,Matsuoka2018,Wang2019,Yang2023}. 
These $z\gtrsim 6$ quasars are powered by intense accretion onto supermassive black holes with masses of up to a few times $10^9$\,\Msol\ \citep{Barth2003,Jiang2007,Kurk2007,DeRosa2011,Yang2020} at or even above
the Eddington limit \citep{Willott2010,DeRosa2011,Zappacosta2023}. These quasars raise the question of how these black holes first formed and how they were able to grow so rapidly.

One theory is that these rare quasars live in the rarest, most massive, dark-matter halos. These halos formed in high-density environments where gas can be efficiently funnelled into galaxies along gas filaments, which could fuel extreme black hole growth.
However, observations have yet to show whether quasars live in the most massive halos, with some surveys searching for protocluster-scale galaxy overdensities around high-$z$ quasars reporting overdensities \citep[e.g.][]{Stiavelli2005,Zheng2006,Morselli2014,Mignoli2020}, while others have found no evidence that quasars reside in higher-density regions \citep[e.g.][]{willott_2005,Banados2013,Simpson2014,Goto2017}. 
This picture has not  gained clarity, not even with the high-quality data delivered by the James Webb Space Telescope \citep[JWST;][]{Gardner2006,Gardner2023,McElwain2023}, as observations have located quasars in a wide range of environments \citep[e.g.][]{Kashino2022,Wang2023,Eilers2024,Champagne2024}.

Some theories suggest that local ({kiloparsec}-scale) interactions can be physical triggers for active galactic nuclei (AGNs) and black hole growth
\citep[e.g.][]{Sanders1988,Hopkins2006}. 
Observations at low-$z$ have indeed suggested that mergers drive at least some AGN activity \citep[e.g.][]{Ellison2011,Bessiere2012,Glikman2015,Araujo2023}. For example, low-$z$ Seyfert galaxies have tidal disturbances in their \ions{H}{i} gas seen at much higher rates than similar inactive galaxies \citep{Lim1999,Kuo2008,Tang2008}. However,
interactions are not ubiquitously observed to be fuelling low-$z$ quasar hosts \citep{Cisternas2011, Kocevski2012, Mechtley2016,Marian2019}. Nevertheless, in the early Universe, where merger rates are higher \citep{Duncan2019,Duan2024a} and black hole growth is more intense, interactions may contribute to a significant amount of black hole growth. Indeed, large fractions of $z>4.5$ AGNs exhibit companion galaxies or merger features \citep[e.g.][]{Duan2024}.
Observations in the rest-frame far-infrared (FIR) with the Atacama Large Millimeter Array (ALMA) have detected companion galaxies around some high-$z$ quasars at ${\lesssim}60$\,kpc separations, interpreted as major galaxy interactions \citep[e.g.][]{Wagg2012,Decarli2017}. In some high-$z$ quasar samples, up to 50\% have such companions \citep{Trakhtenbrot2017}. 
ALMA has also revealed gas bridges connecting quasar hosts and their nearby companions, clearly indicating ongoing interaction \citep[e.g.][]{Izumi2024,Zhu2024}. 
Rest-frame UV observations have also discovered some kiloparsec-scale companions around high-$z$ quasars \citep{McGreer2014,Farina2017,Marshall2019c,Mazzucchelli2019}, albeit less frequently than in the FIR \citep{willott_2005}. This is likely due to the fact that many companion galaxies are dusty \citep{Trakhtenbrot2017}.

The local kiloparsec-scale environments of high-$z$ AGNs are now being uncovered in the rest-frame UV and optical with JWST, with new insights achieved thanks to its higher sensitivity and resolution, combined with less dust attenuation in the optical.
Many companion galaxies have been discovered within ${<}10$\,kpc of $3<z<6$ AGNs \citep{Perna2023,Perna2023b,Matthee2024,Uebler2023,Ji2024}. Merger signatures have also been found around even higher-redshift AGNs: $z\simeq7$ \citep{Merida2025}, $z = 7.15$ \citep{Uebler2024}, and $z=8.7$ \citep{Larson2023}.
In the brighter `quasar' regime, the previously discovered quasar--galaxy merger 
PSO J308.0416–21.2339 at $z\simeq6.2$ \citep{Decarli2019} has been followed up with the JWST/NIRSpec integral field unit \citep[IFU;][]{Boeker2022} to provide a detailed analysis of the gas and stellar emission from the system \citep{Loiacono2024,Decarli2024}. The NIRSpec IFU also discovered that three quasars at $z=6.8$--7.1 are undergoing mergers with nearby companion galaxies \citep{Marshall2023,Marshall2024}.
While companion galaxies and ongoing mergers are not ubiquitous, this growing sample suggests that mergers could be important for driving early black hole growth.

\citet{Mechtley2014} and \citet{Marshall2019c} observed six $z\simeq6$ quasars with deep Hubble Space Telescope (HST) Wide Field Camera 3 (WFC3) infrared (IR) imaging with the aim of detecting their host galaxies. 
To search for the light from the hosts, the quasar emission was modelled and subtracted via detailed subtraction techniques \citep{Mechtley2012,Mechtley2014,psfMC}. The subtracted images revealed numerous galaxies that had been hidden by the quasar emission. The $J$- and $H$-band colours and magnitudes of the neighbouring galaxies suggested that up to nine, surrounding five of the six quasars, could be potential $z\simeq6$ quasar companions \citep{Marshall2019c}. These potential companions are separated from the quasars by $1\farcs4$--$3\farcs2$ (a projected 8.4--19.4~kpc if at the same redshift) and, thus, they could  be interacting with the quasar hosts. 
The companions have UV absolute magnitudes of $-22.1$ to $-19.9$ mag and UV spectral slopes $\beta$ of $-2.0$ to $-0.2$, consistent with luminous star-forming galaxies at $z\simeq6$ \citep{Bouwens2012,Dunlop2012,Finkelstein2012,Jiang2013,Jiang2020}. 
The Prime Extragalactic Areas for Reionization and Lensing Science (PEARLS) Guaranteed-Time Observation (GTO) programme \citep{Windhorst2023} has followed up two of these six 
$z\simeq6$ quasars with NIRSpec IFU observations.
This paper presents the results for NDWFS J142516.3+325409, herein `\NDWFS'. The data for the second quasar SDSS J000552.34–000655.8 will be presented in a later paper.

\NDWFS\ was discovered with the Hectospec spectrograph on the MMT and has a Lyman-$\alpha$ redshift $z=5.85$ \citep{Cool2006}. With the Plateau de Bure Interferometer (PdBI), CO (6--5) emission was detected at $z=5.8918\pm0.0018$ 
\citep{Wang2010}.
\citet{Shen2019} obtained rest-frame UV spectroscopy with GNIRS on Gemini-North,
measuring a black hole mass from \ions{C}{iv} of $2.5\substack{+0.7\\-0.6}\times10^9\,M_\odot$. Combined with a bolometric luminosity of $(9.53\pm0.09)\times10^{46}$ erg\,s$^{-1}$, this result corresponds to an Eddington ratio of $0.29\substack{+0.08\\-0.06}$.

PEARLS targeted \NDWFS\ because it has two companion galaxies potentially at the redshift of the quasar \citep{Mechtley2014,Marshall2019c}.
The Large Binocular Camera (LBC) on the Large Binocular Telescope (LBT) detected no $g$-band flux from these companions to $m_g\gtrsim28.3$\,mag \citep{Marshall2019c}. In the $r$ band, there is only a constraint of $m_r\gtrsim25.7$\,mag for the companion furthest from the quasar on the sky \citep{Mechtley2014}. The $J$- and $H$-band detections combined with faint $g$- and $r$-band limits exclude the possibility that these companions are blue foreground galaxies, but they could be red, luminous galaxies at $z\simeq1.1$.
Further evidence of close companions comes from the \citet{Cool2006} spectrum, which has an absorption feature at ${\sim}8350$\,\AA, ${\sim}20$\,\AA\ red-wards of Lyman-$\alpha$. This could come from \ions{H}{i} absorption from a companion galaxy infalling at ${\sim}720\textrm{~km~s}^{-1}$ \citep{Mechtley2014}. 
Thus, the presence of close companions around \NDWFS\ seems likely but requires spectroscopic confirmation.

This paper presents the NIRSpec IFU observations of \NDWFS. Our goals are to confirm the redshift of the two potential companion galaxies to determine whether this is a merging system, study the ionised-gas properties of the host galaxy and the companions, and measure the quasar black hole mass and accretion rate.
This paper is organised as follows. Section~\ref{sec:observations} describes the observations, data reduction, and analysis. Section~\ref{sec:BHproperties} presents the quasar and black hole properties including the redshift, luminosity, black hole mass, and Eddington ratio. 
Section~\ref{sec:GalaxyResults} focuses on the companion galaxies, presenting a range of properties including their star-formation rates, kinematics, and stellar masses. 
Section~\ref{sec:Lens} presents a potential foreground lensing galaxy discovered in the data.
Section~\ref{sec:discussion} discusses our findings and Section~\ref{sec:Conclusions} gives an overall summary.

Throughout this work, we adopt the WMAP9 cosmology \citep{Hinshaw2013} as included in \textsc{AstroPy} \citep{Astropy2013} with $H_0=69.32$ \kms\,Mpc$^{-1}$, $\Omega_{\rm M}=0.2865$, and $\Omega_\Lambda=0.7134$.
All quoted physical separations are proper distances. At the redshift of the quasar, $z=5.89$, $1\arcsec=5.92$\,kpc in this cosmology. 

\section{Observations and data analysis} 
\label{sec:observations}

\subsection{JWST observations}
JWST observations of \NDWFS\ were taken as part of the PEARLS GTO programme \citep[PI R. Windhorst, PID \#1176,][]{Windhorst2023}.
The quasar was observed with the NIRSpec IFU in the prism+clear configuration, which gives a spectral resolution of $R\sim100$ covering a wavelength range of 0.6--5.3\,\micron.
Wide-aperture target acquisition (WATA) was performed on the quasar to centre the IFU at R.A.\ 14:25:16.4078, Decl.\ +32:54:09.580, slightly offset from the quasar position. This ensured that the extended neighbouring galaxies would be placed within the $3\arcsec\times3\arcsec$ field of view (FoV)\null.
The observations were taken on 2024 Feb~14 at a position angle of the telescope's V3 axis of 271\fdg6.

The science observations consisted of two separate sets, both using the NRSIRS2RAPID readout pattern with 31 groups per integration, one integration, and a four-point dither pattern to give 1867\,s of exposure time. Combining the two sets gave a total science exposure time of 3735\,s. 
Within the four-point dither pattern, each dither has 0\farcs025 sub-pixel offsets that improved the sampling of the point-spread function (PSF)\null.
One `leakcal' observation set was also taken with observation specifications identical to the individual observation sets: a four-point dither pattern with the same readout pattern and 1867\,s total exposure time. The leakcal was performed with the IFU aperture closed in order to measure spectral contamination from failed-open micro-shutter assembly (MSA) shutters. This leakcal was applied to both observation sets.
The observations did not include separate background exposures because the sources are small enough to use blank areas within the FoV for background subtraction.

\subsection{Data reduction}

To reduce the observations, we used the JWST pipeline version 1.14.0 \citep{Bushouse2022} with CRDS context file \textsc{jwst\_1256.pmap}. 
We followed the IFU reduction process provided by the TEMPLATES team \citep{Rigby2023,TEMPLATESgithub}, except for differences noted below.
We included the snowball-masking jump-detection step inbuilt into the JWST pipeline in Stage~1 to remove large cosmic-ray artefacts.
We corrected $1/f$ noise using the NSClean package \citep{NSClean}.
In Stage~2, we subtracted the leakcals from the science exposures dither-by-dither. This improves the quality of the final data cube, giving fewer spurious emission and absorption features; this improvement is more beneficial than the slight increase in background noise introduced by using the leakcals.
We included the default pipeline outlier-detection step in Stage~3, which works well for removing outliers in our data cube.
To combine the data cubes, we used the \textsc{drizzle} algorithm and built the cube in sky coordinates. The output data cube has a pixel scale of 0\farcs05, sub-sampled relative to the detector's 0\farcs1 pixel scale, possible because of the dithering.

As the astrometry for the IFU is slightly uncertain, we assigned a new WCS based on the known quasar position from the Panoramic Survey Telescope \& Rapid Response System (Pan-STARRS), which is astrometrically aligned to Gaia EDR3 \citep{White2022}. We found the peak, central sub-pixel quasar location via image smoothing and assigned this to the Pan-STARRS mean DR2 position, R.A.\ 14:25:16.3286, Decl.\ +32:54:09.554. This required only a simple shift by 
${\lesssim}0\farcs1$
with no rotation. \citet{Jones2024} presented a similar example with further details.

\subsection{Analysis technique}
This section details the techniques for our analysis of the data. The resulting measurements are given and discussed in Sections~\ref{sec:BHproperties} and~\ref{sec:GalaxyResults}.

\subsubsection{Background and continuum subtraction}
\begin{figure*}
\begin{center}
\includegraphics[scale=0.86]{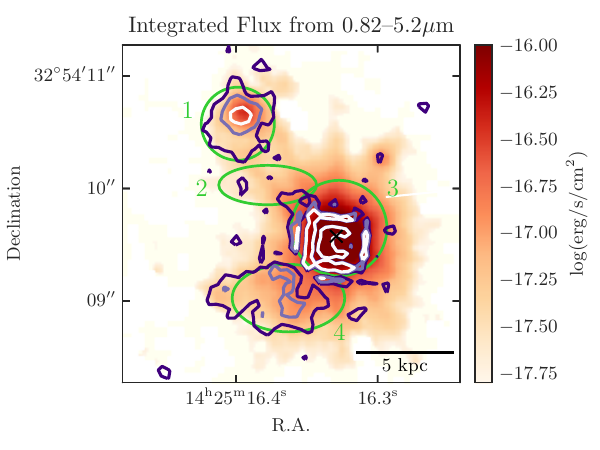}
\includegraphics[scale=0.86]{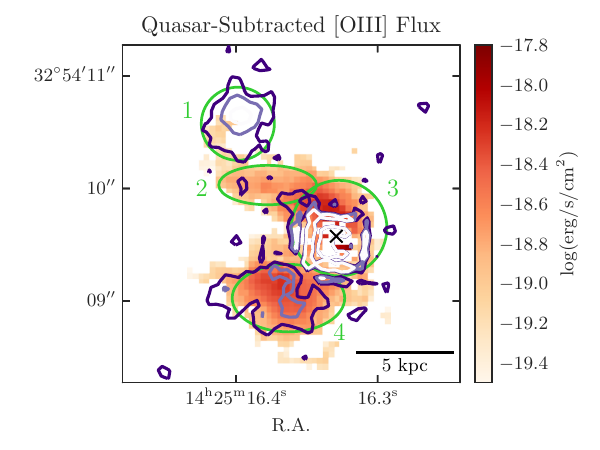}
\caption{Flux from the background-subtracted data cube integrated across 0.82--5.2\micron\ (left) and the \oiiia\ flux map after quasar subtraction (right). The fluxes in each spaxel are as indicated in the respective colour bars. Purple to white contours show the quasar-subtracted HST/WFC3 F125W surface brightness \citep[][who found very similar F160W morphology]{Marshall2019c}. Dark purple, light purple, and white correspond to 24.75, 24.00, and 23.25 mag arcsec$^{-1}$, respectively. Four regions are marked by  green elliptical apertures: region~1 is the north-eastern companion galaxy, region~2 is a bridge of connecting gas, the quasar host-galaxy is region~3, and region~4 is the south-eastern companion galaxy. 
The maps are aligned along the cardinal directions, with north upwards and east to the left.
}
\label{fig:HST}
\end{center}
\end{figure*}

The first analysis step is to subtract the sky background from all spaxels in the data cube. The selected background region is a circular aperture of radius 0\farcs5 near a corner of the IFU FoV, chosen to avoid the extended PSF of the quasar and emission from potential companions as well as the higher noise pixels around the edge of the FoV\null. We defined the background spectrum as the median spectrum across spaxels within this aperture, which we then subtracted from each spaxel.
Figure \ref{fig:HST} shows the flux in this background-subtracted data cube integrated over 0.82--5.2\,\micron. We excluded the wavelengths from 0.6--0.82\micron, blue-wards of Lyman-$\alpha$ at $z=5.8$ because they are dominated by noise.

To simplify fit the emission lines, which is the focus of this work, we subtracted the continuum emission from the data cube.
The continuum emission was measured from flux densities in the emission-line- and iron-continuum-free windows at rest frame: 
1320--1330\,\AA,
1455--1470\,\AA,
1690--1700\,\AA,
2160--2250\,\AA,
3010--3040\,\AA,
3240--3270\,\AA,
3790--3810\,\AA, 
4180--4230\,\AA,
5550--5630\,\AA, and 5970--6000\,\AA. These windows are based on the continuum windows of \citet{Kuraszkiewicz2002} and \citet{Kovacevic2010}, although we slightly increased the width of some of their windows to be more compatible with our low-resolution spectra.
We also added two additional windows near the red edge of the spectral range at 
7080--7130\,\AA\ and 7370--7630\,\AA\ to better constrain the red end of the rest-frame-optical continuum.
We characterise the spaxels into three categories based on their median signal-to-noise ratio, $S/N$, in the continuum windows above 2.6\,\micron---bright continuum with $S/N>4.5$, intermediate continuum with $S/N>3.5$, and faint continuum with $S/N>1.5$.
For the bright-continuum spaxels, we estimated the continuum level by cubic-spline interpolation between the median fluxes of each continuum window.
For the intermediate-continuum spaxels, we used the cubic spline interpolation for $\lambda>2.6\,\micron$, and for $\lambda\leq2.6\,\micron$ we fit a quadratic function to the continuum window medians at $\lambda<3\,\micron$ (with the overlap region in the polynomial fit helping to minimise the discontinuity at $\lambda=2.6\,\micron$).
For the faint continuum, we used two quadratic function fits: for $\lambda\leq2.6\,\micron$, a fit to the window medians at $\lambda<3\,\micron$, and for $\lambda>2.6\,\micron$, a fit to the window medians at $\lambda>2.55\,\micron$. We used quadratic fits for lower $S/N$ spectra because the cubic interpolation can introduce artificial oscillations due to over-fitting of the data. We did not subtract the continuum in spaxels with $S/N<1.5$; any remaining continuum is negligible and varies minimally with wavelength and, therefore, it does not affect the emission-line fitting. 
These continuum models were subtracted from the spectrum spaxel-by-spaxel to create a continuum-subtracted cube.

As found by \citet{Uebler2023}, \citet{DelPino2024}, \citet{Jones2024}, and \citet{Lamperti2024}, the noise estimation from the default `ERR' cube output from the reduction pipeline underestimates the true noise seen in the data cube. Following the approach of those studies, we increased the ERR cube by a constant factor within each spaxel to match the observed noise. 
To estimate these multiplicative factors, we measured the root mean square (RMS) noise across the continuum windows of 5550--5630\,\AA\ and 5970--6000\,\AA, the two windows between \hb\ and \ha, in the continuum-subtracted cube. In spaxels where the ratio of the RMS to the mean ERR in the same windows is greater than 1, we multiplied the ERR spectrum by RMS/ERR to give a more realistic estimation of the uncertainty in the data cube. The median multiplicative factor across the data cube is 1.55.

\subsubsection{Quasar spectral fitting}
\label{sec:QuasarFitting}

\begin{figure*}
\begin{center}
\includegraphics[scale=0.86]{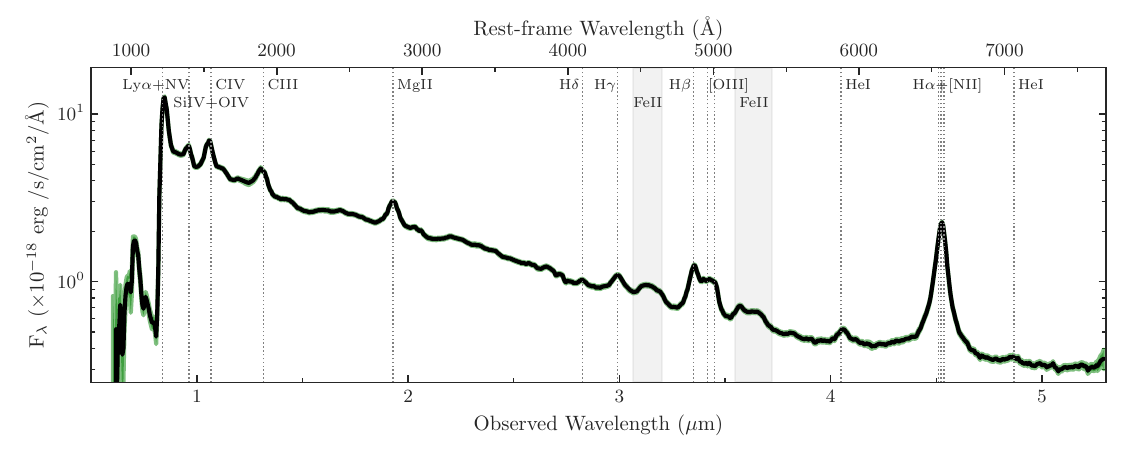}
\caption{Integrated quasar spectrum (black curve) showing the full wavelength range covered by the NIRSpec prism. The spectrum is measured within an aperture of radius 0\farcs35 centred on the quasar emission peak and has an aperture correction of 1.119$\times$ applied. The green region shows the $\pm5\sigma$ uncertainty level of the spectrum, showing that  $S/N\gg5$ for $\lambda>0.65$\,\micron. Key quasar and galaxy emission lines are marked with dotted vertical lines and regions of strong \ions{Fe}{ii} emission, where the numerous transitions form a pseudo-continuum, are shaded.}
\label{fig:QuasarSpectrum}
\end{center}
\end{figure*}

To estimate the black hole properties, we performed a model fitting on the integrated quasar spectrum around \hb\ and \ha.
We integrated the background-subtracted cube and not the continuum-subtracted cube, as the integrated quasar continuum is able to be modelled within our fitting framework.
We integrated the background-subtracted cube across an aperture centred on the peak of the quasar emission.
We chose the aperture radius to be the outer radius of the inner core of the PSF, just prior to the ring of reduced flux; this maximises the quasar $S/N$, while containing the majority of the quasar flux.
These aperture radii, based on the PSFs  described in Section~\ref{sec:quasarSubtraction} and Appendix~\ref{sec:PSF}, are 0\farcs25 for the spectrum around \hb\ at $\sim3.4$\micron\ and 0\farcs35 for the region around \ha\ at $\sim4.5$\micron.
To determine the appropriate aperture corrections, we measured how much of the AGN broad-line flux is contained within these apertures relative to that in the full PSF shape. This gives an aperture correction of 1.149$\times$ for \hb\ and 1.119$\times$ for \ha.   We applied these aperture flux corrections to the integrated quasar spectrum.
Figure~\ref{fig:QuasarSpectrum} shows the 0\farcs35 aperture integrated spectrum.

Using this integrated spectrum, we performed a model fitting using the code \textsc{QubeSpec}\footnote{\url{https://github.com/honzascholtz/Qubespec}}, which models spectra via a Markov chain Monte Carlo (MCMC) process.
We fit each narrow emission line (\hb, \oiii, \ha, and \nii)  with two Gaussian components: one for the true `narrow' galaxy emission and one for a broader `outflow' component.
For the quasar broad-line region (BLR), we fit the broad \hb\ and \ha\ lines each with two Gaussians centred on the same redshift, namely, a double Gaussian model, as used previously for high-$z$ quasars observed with JWST \citep[e.g.][]{Yue2023,Marshall2024}.
This gives a better fit to the spectrum than a single Gaussian BLR model.
We fit the continuum emission as a simple 1D power law across the wavelength region of interest.
We also included a template \citep{Park2022} for the quasar \ions{Fe}{ii} emission.

The \oiii\ and \nii\ doublet flux ratios were fixed to standard values, $2.98$ for \oiiia/\oiiib\ \citep{Storey2000} and $3.05$ for \niia/\niib\ \citep{Dojcinovic2023}.
Galaxy emission lines are expected to have $F_{\rm{H}\alpha}/F_{\rm{H}\beta}=2.86$ for Case B recombination at a temperature of $T = 10^4$\,K and electron density of $n_e = 10^2~\rm{cm}^{-3}$ \citep{Osterbrock1989}. Dust attenuation and an AGN contribution would make the ratio larger than 2.86. Therefore, we constrained the narrow line, outflow, and two BLR \ha\ line peak amplitudes to be between 2.5--10 times larger than those of \hb, to aid the model in finding a physically realistic fit to the data.
We chose a lower limit of 2.5 instead of 2.86 to account for uncertainties and systematic effects. If a system with a true ratio of 2.86 was observed with $S/N=10$ for the fainter \hb\ line, a $1\sigma$ offset in the measured line fluxes would result in a measured ratio of 2.5. Choosing a lower limit of 2.5 allows for such measurement errors.

We constrained the `narrow line' components for \ha\ and \nii\ to  the same redshift and physical velocity width across all three emission lines because we assumed they arise from the same physical region with the same kinematics. We also imposed the same constraint for the broader `outflow' components, independent of the properties of the `narrow' components. The `narrow' and `outflow' components of the \hb\ and \oiii\ lines are constrained in the same way, with this wavelength region fit separately from the \ha--\nii\ region, as discussed below.
The \ions{Fe}{ii} emission is constrained to have the same redshift as the \hb\ BLR in the fit, given the low spectral resolution of our data \citep[][their Fig.~16]{Kovacevic2010}. We constrained the \ions{Fe}{ii} line to have full width at half maximum of $\rm{FWHM}_{\rm{obs}}<2100$\,km\,s$^{-1}$; otherwise, it becomes artificially large and gives a visually poorer fit.

The observed line width is a convolution of the physical velocity width of the line with the instrumental velocity width of $\rm{FWHM}_{\rm{obs}}=\sqrt{\rm{FWHM}_{\rm{phys}}^2+\rm{FWHM}_{\rm{inst}}^2}$\,. Because the NIRSpec prism spectral resolution varies by a factor $\geq$5 from 0.6 to 5.3\,\micron, 
$\rm{FWHM}_{\rm{inst}}$ is significantly larger at the wavelengths of \hb--\Oiii\ than at \ha--\Nii \ and that must be accounted for.
Appendix~\ref{sec:LSF} presents the line spread function (LSF) for the NIRSpec IFU prism measured using observations of the planetary nebula SMP LMC~58. Our measured spectral resolution is ${\sim}14$\% higher than the pre-flight expectations, leading to 
$\rm{FWHM}_{\rm{inst, H}\beta}=2083\pm188$\,km\,s$^{-1}$ and
$\rm{FWHM}_{\rm{inst, H}\alpha}=1126\pm102$ km\,s$^{-1}$. Our \textsc{QubeSpec} fitting uses these two constant values across the \hb--\Oiii\ and \ha--\Nii\ wavelength regions, respectively. The change in resolution across each wavelength range is $\lesssim100$\,\kms, insignificant relative to the ${\sim}200$\,km\,s$^{-1}$ uncertainty (Section~\ref{sec:MapFitting} and Appendix~\ref{sec:LSF}).

We fit the two wavelength regions of \hb--\Oiii\ and \ha--\Nii\ separately because the wavelength variation of the PSF warrants different integration apertures. However, as the emission lines arise from the same physical system, after first fitting the \hb--\Oiii\ region, we can use the resulting fit parameters in the priors for the \ha--\Nii\ fit to ensure that the line velocities and velocity widths are generally consistent.
In particular,
we defined the priors for the narrow line, outflow, and two BLR line redshifts and widths to be normally distributed around the best-estimate redshifts and line widths from the \hb--\Oiii\ fit, accounting for the instrumental resolution.

\begin{figure*}
\begin{center}
\includegraphics[scale=0.86]{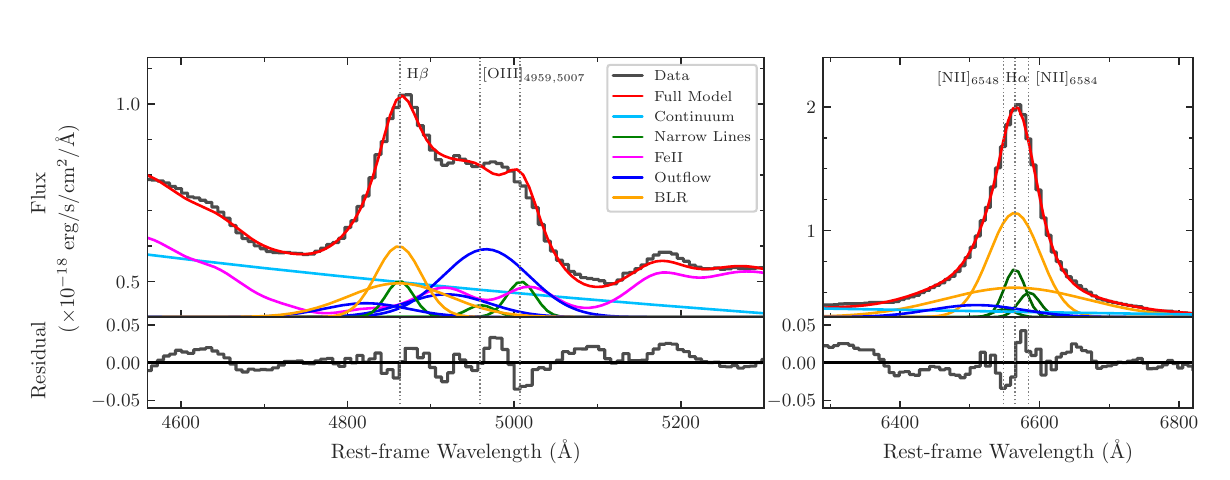}
\caption{Integrated quasar spectrum in an aperture centred on the peak of the quasar emission:\ the region around \hb--\Oiii\ with an aperture radius of 0\farcs25 (left) and the region around \ha--\Nii\ with an aperture radius of 0\farcs35 (right). The solid black line in the upper panel shows the observations, and the red line shows the best model fit from \textsc{QubeSpec}.
The individual components of the model fit are also shown as indicated in the legend: the power-law continuum emission, Gaussian narrow and outflow lines, \ions{Fe}{ii} template \citep{Park2022}, and double-Gaussian BLR model.
The emission-line model components are offset by $0.4\times10^{-18}$\,erg\,s$^{-1}$\,cm$^{-2}$\,\AA$^{-1}$ in the left panel and $0.3\times10^{-18}$erg\,s$^{-1}$\,cm$^{-2}$\,\AA$^{-1}$ in the right panel to aid in visualisation.
The lower panels show the residual between the observed flux and the full model.
No aperture correction is applied to this figure, but an aperture correction of 1.119 $\times$ for \ha\ and 1.149 $\times$ for \hb\ is applied to all calculations from these spectral fits.
}
\label{fig:QuasarFitting}
\end{center}
\end{figure*}

We ran the \textsc{QubeSpec} MCMC for 80\,000 iterations. \textsc{QubeSpec} uses 64 walkers
with a burn-in of 25\% of the iterations.
The resulting model fit is shown in Figure~\ref{fig:QuasarFitting}.
To determine the resulting emission-line properties, we randomly chose 100 sets of model-parameter values from the output chain; this adequately samples the full distribution. We then calculated the 16th, 50th, and 84th percentiles of the resulting line fluxes, luminosities, and FWHMs (Section~\ref{sec:BHproperties}). 

\subsubsection{Quasar subtraction}
\label{sec:quasarSubtraction}
Measuring the underlying emission from the host galaxy requires modelling and subtracting the quasar emission from the data cube. The quasar emits both narrow emission lines from the narrow line region (NLR) and broad emission lines from the BLR, which is a spatially unresolved point source at this redshift. While the host galaxy also emits narrow emission lines through star formation, only the quasar BLR produces the broad emission features seen in quasar spectra. Thus a map of the BLR flux is a map of the quasar's point-source emission, that is, the instrumental PSF\null. 
We can use this PSF to subtract both the quasar BLR and spatially unresolved NLR emission from the data cube, to isolate the host galaxy emission.

Using the software \textrm{QDeblend3D} \citep{Husemann2013,Husemann2014}, we used the BLR flux in each spaxel to determine the PSF shape.  We measured the flux in the BLR line wings, avoiding the line centres where the narrow lines would contaminate the measurement.
We measured these BLR fluxes for both the \hb\ and \ha\ broad lines because the PSF varies with wavelength.
For \ha, we measured the flux across rest-frame 6485--6525\,\AA\ and 6605--6640\,\AA, either side of the emission peak at 6564.6\,\AA.
For \hb, we measured the flux across rest-frame 4800--4850\,\AA, on the blue side of the peak at 4862.68\,\AA; we only considered the blue wing due to the blending with \oiiib\ on the red-ward side of the line.
Figure~\ref{fig:PSFs} shows the resulting 2D \ha\ and \hb\ BLR flux maps, that is, the IFU PSF shape at 3.35 and 4.52\,\micron.

We then took the spectrum from the central, brightest spaxel
as our estimated quasar spectrum. This assumes that
the quasar flux is dominant and 
the host flux is negligible, a reasonable assumption across much of the spectral range given the non-detection of the host flux in the PSF-subtracted HST F125W and F160W images \citep{Marshall2019c}.
Any host contribution 
in this spaxel 
will lead to a mild oversubtraction, however, it will be insignificant given the relative fluxes of the quasar and host.
After scaling the quasar spectrum by the 2D PSF to create a quasar cube, we subtracted the quasar cube from the original continuum-subtracted data cube, resulting in a galaxy emission cube where the quasar contribution has been removed.

The quasar subtraction removes all spatially unresolved emission centred at the location of the BLR (e.g. the quasar's unresolved NLR flux and any spatially unresolved outflows).
If spatially resolved outflow features are present within the BLR line wings, this would result in a poor measurement of the PSF and a poor quasar subtraction, with significant negative and positive residuals around the BLR core and the side of the line opposite the outflow.
From the residual spectra and the resulting
measured PSFs (Figure~\ref{fig:PSFs}), we are confident that our subtraction works well and is not significantly affected by resolved outflows.

The quasar subtraction adds additional uncertainty to the resulting data cube. To estimate this increase \citep[following][]{Marshall2023,Marshall2024}, we generated 200 realisations of the continuum-subtracted cube, applying normally distributed noise from our rescaled `ERR' array and 
performing the same quasar subtraction on each realisation. We calculated the standard deviation across the resulting 200 host-galaxy cubes at each wavelength in each spaxel.
We determined the ratio of the mean standard deviation around the peaks of \ha\ and \hb\ within rest-frame 25\,\AA\, to the same measurement on the original host cube, and multiplied the rescaled `ERR' array by these fractions. The maximum uncertainty increase around \hb\ is a factor of 2.17, and the maximum around \ha\ is a factor of 3.02.

Because we performed the quasar subtraction separately for \hb\ and \ha, we use the \hb-subtracted data cube and ERR cube for $\lambda\leq4\,\micron$ and the \ha-subtracted data cube and ERR cube for $\lambda>4\,\micron$. However, the PSF varies significantly with wavelength. This paper considers only lines in the immediate vicinity of \hb\ and \ha, where no significant PSF-variation effects are expected. At wavelengths away from these lines, the quasar-subtracted data will be inaccurate, and a different method of quasar subtraction would be required.

\subsubsection{Host line fitting}
\label{sec:MapFitting}
We mapped the extended emission within both the full (non-quasar subtracted) and quasar-subtracted (host-galaxy) data cubes to visualise the effect of the quasar subtraction.
We only considered the emission lines immediately surrounding \hb\ and \ha, as the quasar subtraction is only reliable at these wavelengths.

We modelled the emission lines at each spatial position with Gaussian functions. The quasar also produces \ions{Fe}{ii} emission, which we modelled using an iron emission template from \citet{Park2022} in the full, non-quasar-subtracted data cube. For spaxels in the full data cube where the mean $S/N$ across rest-frame 5150--5370\,\AA\ (in the \ions{Fe}{ii} emission region) is greater than 2, we fit the `quasar spectral model', which models the \ha\ and \hb\ lines with
two separate Gaussian components, one for the narrow galactic emission and a broader component for the BLR emission, the \oiii,\ and \nii\ lines as single narrow Gaussians, and includes the \citet{Park2022} \ions{Fe}{ii} template. In other spaxels in the full (non-quasar-subtracted) cube and in all spaxels within the quasar-subtracted cube, we fit a single narrow Gaussian function for each of \hb, \oiii, \nii, and \ha. These fits include no \ions{Fe}{ii}.

The fits constrained the \Oiii\ and \Nii\ doublet flux ratios to $2.98$ for \oiiia/\oiiib\ and $3.05$ for \niia/\niib.
We constrained all of the narrow lines to have the same central velocity and physical width, $\rm{FWHM}_{\rm{phys}}$.
The instrumental velocity widths (Appendix~\ref{sec:LSF}) are
$\rm{FWHM}_{\rm{inst}}=2083\pm188$\,km\,s$^{-1}$, $1997\pm181$\,km\,s$^{-1}$, $1966\pm178$\,km\,s$^{-1}$, and $1126\pm102$\,km\,s$^{-1}$ at the wavelengths of \hb, \oiii, and \ha\,, respectively, assuming that $\rm{FWHM}_{\rm{inst, [NII]}}=\rm{FWHM}_{\rm{inst, H}\alpha}$.
We corrected the observed $\rm{FWHM}_{\rm{obs}}$ using these $\rm{FWHM}_{\rm{inst}}$ to constrain each line fit to have the same $\rm{FWHM}_{\rm{phys}}$.
The broad \hb\ and \ha\ components were similarly constrained to have the same central velocity and physical velocity width as each other (but different with respect to the narrow lines).

To produce the line maps, the line velocity and width and the integrated flux were measured from the Gaussian fit to the lines in each spaxel. The velocity was measured as the Gaussian line peak relative to the quasar host redshift of $z=5.901$. We corrected the measured Gaussian FWHMs for the instrumental resolution, that is, the maps depict $\rm{FWHM}_{\rm{phys}}$.
The maps include spaxels where the emission line $S/N>3$ and also any adjacent spaxel with $S/N>1.5$, following the \textsc{find\_signal\_in\_cube} algorithm of \citet{SunGithub}.

Figure \ref{fig:HST} shows the \oiiia\ flux map after quasar subtraction. 
The quasar-subtracted HST/WFC3 F125W image from \citet{Marshall2019c} is overlaid for comparison, with the F160W image showing very similar morphology. 
The integrated 0.82--5.2\micron\ flux map, \oiiia\ flux map, and HST images show four clear structures in this system, depicted by the ellipses in Figure~\ref{fig:HST} and labelled regions 1--4 from north to south.
Region~1 lies to the north-east of the quasar, where faint \oiiia\ emission is coincident with bright integrated flux.
Region~2 is an area of \oiiia\ emission between region 1 and the quasar host galaxy (region 3) that was not detected in the F125W or F160W HST imaging. 
Region~4 lies south-east of the quasar, emitting both bright \oiiia\ and continuum flux.
Section~\ref{sec:GalaxyResults} discusses these regions in detail.

\subsubsection{Stellar population modelling}
\label{sec:SPSmodelling}
The emission-line maps reveal multiple emission-line regions surrounding the quasar (Figure~\ref{fig:HST}; discussion in Section~\ref{sec:GalaxyResults}).
To estimate the stellar properties of these regions, we performed simultaneous line and continuum fitting with \textsc{pPXF} \citep{Cappellari2004,Cappellari2017,Cappellari2023}, a penalised pixel-fitting algorithm. \textsc{pPXF} estimates the stellar-population properties via stellar population synthesis (SPS) modelling of the spectrum. 
We do not use the emission-line properties as fit by \textsc{pPXF}, instead favouring those from the continuum- and quasar-subtracted line fitting reported in Section~\ref{sec:MapFitting}. We report \textsc{pPXF} results only for the stellar-population properties.

For \textsc{pPXF} to model the stellar population, it must be supplied with input spectra that include the continuum emission. We therefore cannot use the quasar-subtracted data cubes, which are first continuum subtracted (Section \ref{sec:quasarSubtraction}). 
The significant PSF variation with wavelength makes it impossible to accurately subtract the quasar emission across the full wavelength range, outside of the immediate area of \hb\ and \ha, using the quasar-subtraction methods used here. Therefore we cannot create quasar-subtracted cubes that accurately depict the continuum emission.
Instead, we used the original data cube and selected only spatial regions that include negligible quasar flux.
As the region~2 and~4 apertures used throughout the rest of this work overlap with the quasar \hb\ and \ha\ PSF shapes (thus, they contain significant quasar flux), we defined two smaller elliptical aperture regions (regions~2* and~4*, Figure~\ref{fig:PSFs}) that avoid such overlap.
Regions~2* and~4* do not contain all of the flux from the respective sources, resulting in a lower $S/N$, but they show the uncontaminated continuum shape, thus making it possible to  model the stellar populations.
We therefore integrated the spectra within the region~1, 2*, and~4* apertures (Figure~\ref{fig:PSFs}) to obtain a spectrum for each region. The continuum flux in region~2* is very faint with typical $S/N\lesssim5$. To allow for a measurement of the continuum in this region, we rebinned its spectrum onto a $10\times$-coarser wavelength grid. 
Region~3, the host galaxy, is hidden behind the quasar emission and cannot be modelled with the methods used here.

The \textsc{pPXF} modelling of regions~1, 2*, and~4* followed a method similar to that of \citet{Cameron2023} and the \textsc{pPXF} examples.\footnote{\url{https://github.com/micappe/ppxf_examples}} We used the SPS templates from the flexible SPS (FSPS) model \citep[][version 3.2]{Conroy2009,Conroy2010} and a \citet{Salpeter1955} initial mass function (IMF)\null.
We restricted the age of the templates to the age of the Universe at $z=5.89$, 967~Myr.
We adopted the dust attenuation law from \citet{Calzetti2000} and
constrained the dust attenuation to be $0\leq A_v\leq1.5$\,mag.
For the emission lines, we fit a series of single Gaussian functions. 
We divided the emission lines into three groups, with lines in each group constrained to have the same radial velocity and width: hydrogen Balmer lines, UV lines ($\lambda_{\rm{rest}}<3000$\,\AA), and optical lines ($\lambda_{\rm{rest}}>3000$\,\AA). The stellar component was left free to have its own velocity and line width.
To account for the instrumental line width, the templates were smoothed by the instrumental LSF (Appendix~\ref{sec:LSF})\null.
The measured flux blue-ward of Lyman-alpha was set to zero to assist with the fitting.

To estimate the uncertainties on the output properties, we performed bootstrapping following \citet{Kacharov2018} and using 100 samples. Our quoted values and uncertainties are the median, 16th, and 84th percentile values from these 100 samples. The output distributions of $[M/H]$ are not centrally peaked but instead cover a wide range of values and typically peak at one of the edges of the allowed parameter range. In these cases, we quote the minimum or maximum value from the 100 samples as a lower or upper limit. 
All reported stellar-population parameters are flux-weighted values.

\section{Quasar and black hole properties}
\label{sec:BHproperties}

Figure \ref{fig:QuasarSpectrum} shows the full integrated quasar spectrum from 0.6--5.3\,\micron. The spectrum has $S/N>5$ between the Lyman limit at rest-frame 912\,\AA\ and Lyman-$\alpha$ at 1216\,\AA\ but no significant flux blue-wards of the Lyman limit. The spectrum has $S/N>150$ between Lyman-$\alpha$ and observed $5$\,\micron\ (rest $1216<\lambda<7250$\,\AA). There are clear detections of the expected AGN emission lines from Lyman-$\alpha$ to \ha. \ions{S}{ii} is not detected. While \ions{N}{ii} is likely present, it is blended with the \ha\ line and 
therefore difficult
to decompose.

Figure~\ref{fig:QuasarFitting} shows our best-fitting quasar spectral model around \hb--\Oiii\ and \ha--\Nii\ from \textsc{QubeSpec}.
This model describes both regions well with low residuals across the wavelength region. 
However, at this low spectral resolution the emission lines and their various kinematic components are blended together and so, precise modelling and line decomposition is limited. Higher spectral resolution data would allow for a more accurate modelling of the quasar.

\subsection{Redshift and luminosity}

\begin{table}
\caption{Redshift measurements}
\begin{center}
\begin{footnotesize}
\begin{tabular}{lccl}
\hline
Measurement & $z$ & Unc. & Reference\\
\hline
\multicolumn{4}{l}{Quasar and host galaxy}\\
Lyman-$\alpha$ & 5.85\0 & 0.02\0 & \citet{Cool2006}\\
CO & \05.8918 & \00.0018 & \citet{Wang2010}\\
\ions{C}{iii}] & 5.861 & 0.007 & \citet{Shen2019}\\
\ions{C}{iv} & 5.851 & 0.010 & \citet{Shen2019}\\
\ions{Si}{iv} & 5.865 & 0.015 & \citet{Shen2019}\\
Quasar NLR  & 5.890 & 0.010 & This work\\
Quasar BLR & 5.888 & 0.010 & This work\\
Host galaxy & 5.901 & 0.010 & This work\\
\hline
\multicolumn{4}{l}{Companions}\\
NE (region 1) & 5.882\rlap{\tablefootmark{a}} & 0.010 & This work\\
Bridge (region 2) &5.891\rlap{\tablefootmark{a}} & 0.010 & This work\\
SE (region 4) &5.902\rlap{\tablefootmark{a}} & 0.010 & This work\\
\hline
\multicolumn{4}{l}{Possible lensing galaxy}\\
\Oiii, \ha & 1.135 & 0.010 & This work\\
\hline
\end{tabular}

\end{footnotesize}
\end{center}
\tablefoot{The quasar redshifts from this work are from the \hb\ line for the BLR and from the kinematically tied \oiii\ and \hb\ lines for the NLR\null. The host-galaxy and companion redshifts are based on fits to the \oiii, \hb, \nii, and \ha\ narrow lines, which are all kinematically tied to have the same redshift. 
}
\tablefoottext{a}{\footnotesize Redshift measurements of these regions with the cross-correlation programme {\sc xcsao} \citep[][spiral template covering rest wavelengths 2984 to 7379~\AA]{Kurtz1998} are consistent with the tabulated values. When emission lines are excluded from the fit, the $S/N$ of absorption features are too low to give a redshift measurement.
}
\label{tab:Redshifts}
\end{table}

Table~\ref{tab:Redshifts} summarises the available redshift measurements.
The NIRSpec prism can have systematic wavelength offsets $\Delta z\lesssim0.01$ from the more precise grating redshifts \citep[][Perna et al. in prep.]{Bunker2024,PerezGonzalez2024}. Thus, we treated this as the overall redshift uncertainty, $\Delta z = 0.01$, for comparison to other data sets. 
We assume that relative redshift measurements within our data cube are accurate to within half of a wavelength element, $\lesssim25$\,\AA\ or $\Delta z\lesssim0.005$; thus, our redshifts are quoted to a three-digit precision to enable relative redshift comparisons at this level.
The MCMC fits give a quasar redshift $z=5.890\pm0.010$ from the NLR and $z=5.888\pm0.010$ from the BLR\null.
From the integrated quasar-subtracted spectrum, the host galaxy (region~3) has a redshift $z=5.901\pm0.010$ from the kinematically tied \hb, \oiii, \ha, and \nii\ emission lines. 
This host-galaxy redshift is the most relevant for tracing the kinematics throughout the system
and so, we adopted 
$z=5.901\pm0.010$
as the baseline redshift 
for all relative-velocity calculations.

Our redshift estimates are larger than the original redshift estimate from the break in the Lyman-$\alpha$ profile, $z=5.85$ \citep{Cool2006}, as well as those from \citet{Shen2019}, listed in Table~\ref{tab:Redshifts}.
These rest-frame UV emission lines can have velocity offsets up to ${\sim}1000$\,\kms\ from the systemic redshift due to strong internal motions or winds in the BLR \citep[e.g.][]{Eilers2020}, that may explain this discrepancy.
In contrast, our redshift measurement is consistent with the CO redshift of $z=5.8918\pm0.0018$ \citep{Wang2010}, which is a more reliable indicator of the systemic velocity.

We calculated the integrated flux and FWHM of the broad-line components of \ha\ and \hb\ using 100 randomly selected sets of spectral-model parameter values from the MCMC output chain. 
This gives a \ha\ BLR luminosity of 
$42.9\substack{+0.1\\-0.1}\times10^{43}$\,erg\,s$^{-1}$ with  FWHM$_{\rm{phys}}=4132\substack{+194\\-176}$\,km\,s$^{-1}$ and a \hb\ BLR luminosity of 
$8.9\substack{+0.3\\-0.2}\times10^{43}$\,erg\,s$^{-1}$ with FWHM$_{\rm{phys}}=3971\substack{+363\\-440}$\,km\,s$^{-1}$.
The flux ratio $F_{\rm{BLR,H}\alpha}/F_{\rm{BLR,H}\beta}=4.8\pm0.1$ is larger than that for a sample of blue AGNs that are free of dust extinction \citep{Dong2008}, $F_{\rm{BLR,H}\alpha}/F_{\rm{BLR,H}\beta}=3.06\substack{+0.38\\-0.33}$, suggesting that \NDWFS\ has some dust attenuation of the BLR\null. The H$\gamma$ to \hb\ flux ratio gives a consistent amount of dust attenuation. However, because the intrinsic ratio $F_{\rm{BLR,H}\alpha}/F_{\rm{BLR,H}\beta}$ varies among AGNs, we cannot precisely estimate the dust attenuation for single sources, only for statistical AGN samples \citep{Dong2008}, and so we do not correct the BLR properties for dust attenuation.

We measured the 5100\,\AA\ rest-frame continuum flux as $(5.66\pm0.03)\times10^{-19}$\,erg\,s$^{-1}$\,cm$^{-2}$\,\AA$^{-1}$ from the mean of the flux between 5090--5110\,\AA.
To estimate the bolometric luminosity $L_{\rm{bol}}$, we used the 5100\,\AA\ bolometric correction of \citet{Runnoe2012,Runnoe2012err}, $L_{\rm{bol}}\simeq0.75L_{\rm{iso}}$, with
$\log(L_{\rm{iso}}) = (1.017\pm0.001) \log(5100 L_{5100})- (0.088\pm0.034) \alpha_{\lambda,\rm{opt}}$. Here $\alpha_{\lambda,\rm{opt}}$ is the optical slope such that $F_\lambda \propto \lambda^{\alpha_{\lambda,\rm{opt}}}$. We fit our integrated spectrum across the region between rest-frame 3981\,\AA\ and 6310\,\AA, avoiding the \ions{Fe}{ii} emission and \hb\ and \oiii\ emission lines, and measured 
$\alpha_{\lambda,\rm{opt}}=-2.24$.
This gives
$L_{\rm{bol}}=(5.7\substack{+1.3\\-1.0})\times10^{46}$ erg\,s$^{-1}$ for \NDWFS. This is lower than the \citet{Shen2019} estimate of $L_{\rm{bol}}=(9.53\pm0.09)\times10^{46}$ erg\,s$^{-1}$ based on the 3000\,\AA\ luminosity, although the two measurements are consistent within $3\sigma$.

\begin{figure*}
\begin{center}
\includegraphics[scale=0.86]{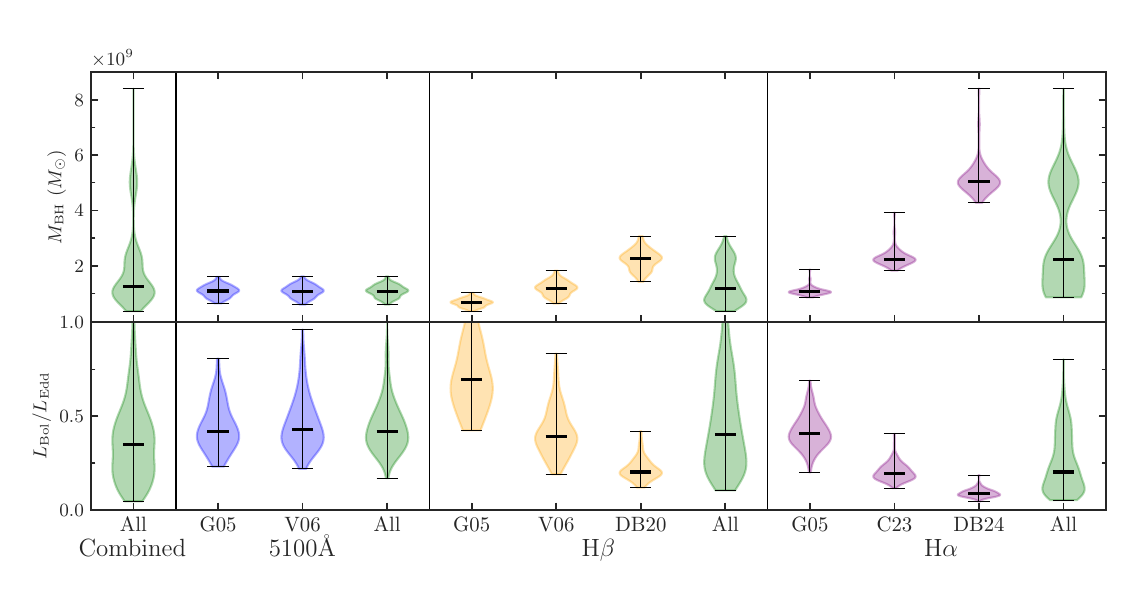}
\caption{Black hole mass and Eddington ratio estimates for \NDWFS\ from single-epoch mass-scaling relations (Equation \ref{eq:MBH}) using the 5100\,\AA\ continuum luminosity and \hb\ line FWHM (blue; second-left panel), the \hb\ line luminosity and FWHM (yellow; middle panel), and the \ha\ line luminosity and FWHM (purple; right panel). Each distribution shows estimates using various calibrations to these relations: \citet[][G05]{Greene2005}, \citet[][V06]{Vestergaard2006}, \citet[][DB20]{DallaBonta2020}, \citet[][DB24]{DallaBonta2024} and \citet[C23]{Cho2023}. The green curves show the combined probability distributions across each of the calibrations from each equation, and from all equations combined (left-most panel), giving each calibration and equation  equal weighting. The horizontal black lines mark the minimum, median, and maximum for each distribution.
The probability distribution reflects the posterior distribution from the MCMC fit, i.e. only the uncertainty in the fitting parameters is included and not the ${\sim}0.4$\,dex scatter from the scaling-relation conversion to a black hole mass.
The top panels show the black hole mass estimates, and the bottom panels show the resulting Eddington ratios.
}
\label{fig:MBH}
\end{center}
\end{figure*}

\subsection{Black hole mass and Eddington ratio}

To estimate the black hole mass of \NDWFS, we used single-epoch mass-scaling relations, which follow the form:
\begin{equation}
\label{eq:MBH}
M_{\rm{BH}} = a ~\left( \frac{L}{L_0} \right)^{b} \left( \frac{\rm{FWHM}}{10^{3}~\rm{km~ s}^{-1}} \right)^{c} \rm{M}_\odot.\quad
\end{equation}
The constants $a$, $b$, and $c$ are calibrated from reverberation-mapping studies at low-$z$ to measurements of the luminosity and FWHM of quasar BLR emission lines, as well as their continuum luminosity \citep[e.g.][]{Wandel1999,Kaspi2000}. We considered three different scaling relation types from the literature, using the 5100\,\AA\ continuum luminosity and the luminosity, $L$, and line FWHM from the \hb\ and \ha\ BLR lines as follows.
\\
\textit{5100\,\AA--\hb\ relation:}
these equations use the 5100\,\AA\ continuum luminosity, with $L/L_0 = \lambda L_{\rm{5100}}/10^{44}$\,erg\,s$^{-1}$, and the FWHM of the \hb\ BLR component.
These relations set $c=2$.
\citet{Greene2005} calibrated the other constants to be $a=(4.4\pm0.2) \times10^6$, $b=0.64\pm0.02$, whereas \citet{Vestergaard2006} derived $a=(8.1\pm0.4)\times10^6$, $b=0.50\pm0.06$.
\\
\textit{Pure \hb\ relation:}
to avoid the difficulty in measuring the continuum luminosity, these equations use the line luminosity of the \hb\ BLR line component, with $L/L_0 = L_{\rm{H\beta}}/10^{42}$\,erg\,s$^{-1}$, and the \hb\ FWHM\null.
\citet{Greene2005} set $c=2$ and derived $a=(3.6~\pm~0.2) \times 10^6$, $b=0.56~\pm~0.02$. \citet{Vestergaard2006}, also with $c=2$, derived $a=(4.7~\pm~0.3)\times 10^6$, $b=0.63~\pm~0.06$.
\citet{DallaBonta2020} found $a=10.1\times~ 10^6$, $b=0.784$, and $c=1.387$.
\\
\textit{Pure \ha\ relation:}
similar to the pure \hb\ relations, these equations use the line luminosity of the \ha\ BLR line component, with $L/L_0 = L_{\rm{H\alpha}}/10^{42}$\,erg\,s$^{-1}$, and the \ha\ FWHM\null.
From \citet{Greene2005}, $a = 2.0~\substack{+0.4 \\-0.3} ~\times~ 10^6$, $b=0.55\pm0.02$, and $c=2.06\pm0.06$.
\citet{Cho2023} found $a=(3.20\pm0.3)\times~ 10^6$, $b=0.61\pm0.04$, with $c=2$.
\citet{DallaBonta2024} found $a=3.58\times~ 10^6$, $b=0.812$, and $c=1.634$.

Using our measured \hb\ and \ha\ broad line properties, we calculated the black hole masses using Equation~\ref{eq:MBH}, for each of the three scaling relation types and each of the quoted calibrations, for the 100 sets of spectral-model parameter values.
Figure~\ref{fig:MBH} shows the distributions of black hole masses for each calibration of each scaling relation. The figure also shows the various calibrations combined, assuming an equal weighting.
Combining all of the scaling relations and calibrations gives
$M_{\rm{BH}}=(1.2\substack{+1.3\\-0.4})\times10^9\,M_\odot$;
however, this estimate does not take into account that each scaling relation has an intrinsic uncertainty of ${\sim}$0.4~dex \citep{Vestergaard2006,DallaBonta2020,DallaBonta2024}. 
Thus, we applied 1000 realisations of $0.4$\,dex noise to the black hole masses calculated from the 100 sets of model parameter values for each scaling relation. As these measurements have non-Gaussian distributions, this is more robust than simply adding the uncertainties in quadrature.
This gives $M_{\rm{BH}}=(1.1\substack{+1.6\\-0.6})\times10^9\,M_\odot$ for the 5100\,\AA--\hb\ relation,
$M_{\rm{BH}}=(1.2\substack{+2.3\\-0.8})\times10^9\,M_\odot$ for the pure \hb\ relation, and $M_{\rm{BH}}=(2.3\substack{+4.9\\-1.6})\times10^9\,M_\odot$
for the pure \ha\ relation. 
While the different methods produce different mass estimates, all estimates are consistent within the large uncertainties.
Combining all three scaling relations and their uncertainties gives 
$M_{\rm{BH}}=(1.4\substack{+3.1\\-1.0})\times10^9\,M_\odot$.
This is consistent with the black hole mass estimate of $(2.5\substack{+0.7\\-0.6})\times10^9\,M_\odot$ from \ions{C}{iv} \citep[][whose uncertainties do not account for the intrinsic scatter in the scaling relations]{Shen2019}.

The Eddington luminosity can be estimated as
\begin{equation}
\begin{split}
L_{\rm{Edd}} &= \frac{4\pi G m_{\rm p}c M_{\rm{BH}}}{\sigma_{\rm T}}= 1.26\times10^{38} \left(\frac{M_{\rm{BH}}}{\rm{M}_\odot}\right) \rm{erg~s^{-1}},\quad
\end{split}
\end{equation}
where $G$ is the gravitational constant, $m_{\rm p}$ the proton mass, $c$ the speed of light, and $\sigma_{\rm T}$ the Thomson-scattering cross-section. We calculated the Eddington ratio, $L_{\rm{Bol}}/L_{\rm{Edd}}$, using each of our black hole mass estimates (Figure~\ref{fig:MBH} shows the results). 
Combining all of the scaling relations and calibrations together gives $L_{\rm{Bol}}/L_{\rm{Edd}}=0.35\substack{+0.22\\-0.20}$, consistent with the \citet{Shen2019} measurement of $0.29\substack{+0.08\\-0.06}$ from \ions{C}{iv}\null.
These uncertainties consider only the measurement uncertainties and not the 0.4~dex intrinsic scatter in the black hole mass scaling relations.
Adding that scatter gives
$L_{\rm{Bol}}/L_{\rm{Edd}}=0.3\substack{+0.6\\-0.2}$. 
\NDWFS\ is most likely accreting at a sub-Eddington rate, typical of quasars at both low- and high-$z$ \citep{Shen2011,Shen2019,Farina2022}, but much higher than the accretion rate for local radio galaxies \citep[e.g.][]{Stasinska2025}.

\subsection{Quasar-driven outflow}

\begin{table*}
\caption{Properties of the quasar-driven outflow measured from \hb\ and \oiiia\ in the integrated quasar spectrum (Figure \ref{fig:QuasarFitting}): luminosity, $L_{\rm{out}}$, line width, $\rm{FWHM}_{\rm{out}}$, velocity offset, $v_{\rm{offset}}$, outflow velocity, $v_{\rm{out}}$, outflow mass, $M_{\rm{out}}$, outflow rate, $\dot{M}_{\rm{out}}$, and power, $P_{\rm{out}}$.
}
\begin{center}
\begin{footnotesize}
\begin{tabular}{lllllllllll}
\hline 
$L_{\rm{out,H\beta}}$\hspace{-0.2cm} & $L_{\rm{out,[OIII]}}$ & $\rm{FWHM}_{\rm{out}}$ & $v_{\rm{offset}}$ & $v_{\rm{out}}$ &  $M_{\rm{out,H\beta}}$\hspace{-0.2cm} & $M_{\rm{out,[OIII]}}$ & $\dot{M}_{\rm{out,H\beta}}$\hspace{-0.2cm} & $\dot{M}_{\rm{out,[OIII]}}$& $P_{\rm{out,H\beta}}$\hspace{-0.2cm} & $P_{\rm{out,[OIII]}}$\\

\multicolumn{2}{c}{\tiny{($10^{43}$ erg / s)}} &  \multicolumn{3}{c}{\tiny{(km / s)}}  & \multicolumn{2}{c}{\tiny{($10^{7}$\,\Msol)}} & \multicolumn{2}{c}{\tiny{($M_\odot$ / yr)}} &\multicolumn{2}{c}{\tiny{($10^{43}$ erg / s)}}\hspace{-0.2cm}\\
\hline 
\vspace{0.1cm}
$1.6 \substack{ +0.0\\-1.2 }$ & 
      $7.9 \substack{ +0.2\\-0.6 }$ & 
      $7080\substack{ +238\\-1186 }$ &
      $-2510\substack{ +450\\-220 }$ & 
      $6050\substack{ +470\\-630 }$ & 
      $27\substack{ +1\\-20 }$ & 
      $6.4\substack{ +0.1\\-0.4 }$ & 
      $1655\substack{ +132\\-1230 }$ & 
      $393\substack{ +31\\-50 }$ &
      $19\substack{ +3\\-14 }$ & 
      $4.5\substack{ +0.6\\-0.9 }$
      \\
\hline
\end{tabular} 
\label{tab:Outflows}
\end{footnotesize}
\end{center}
\tablefoot{These calculations assume a gas density in the ionised outflow of $\langle n_e\rangle = 500\,\rm{cm}^{-3}$, and an outflow radius of $R_{\rm{out}}=1\pm0.1$ kpc, as the outflow is spatially unresolved in the current data.}
\end{table*}

There is a clear detection of an outflow signature from the integrated quasar spectrum in Figure \ref{fig:QuasarFitting}, as traced by a broad kinematic component in \oiii, \hb, and \ha. 
This broad outflow is modelled (Section \ref{sec:QuasarFitting}) with a single Gaussian component for each emission line, with all components constrained to have consistent physical line widths and velocities.
Table \ref{tab:Outflows} lists the outflow properties measured from the \textsc{QubeSpec} MCMC fit to the spectrum.
The \oiiia\ outflow component has $\rm{FWHM}_{\rm{phys}}=7080\substack{+238\\-1186}$\,km\,s$^{-1}$ (corrected for instrumental resolution)
with a velocity offset of $-2510\substack{+450\\-220}$\,km\,s$^{-1}$ relative to the narrow Gaussian component.
This corresponds to an outflow velocity $v_{\rm{out}}=v_{\rm{offset}}+{\rm{FWHM}}/2=6050\substack{+470\\-630}$\,km\,s$^{-1}$.
This high-speed outflow is comparable to those of the most extreme quasar-driven outflows in the Universe \citep[e.g.][]{Perrotta2019,Zamora2024,Vayner2024}.
This velocity is much larger than the most extreme starburst-driven outflows \citep{Heckman2016,Perrotta2021} and the velocities caused by galaxy interactions and mergers \citep{Soto2012,Rich2015}, implying that the outflow must be AGN-driven.
Potential mechanisms to explain such an extreme AGN-driven outflow may be ultra-fast nuclear winds, which shock and accelerate the ISM, or acceleration by radiation pressure on dust grains (see \citealt{Zakamska2016} and \citealt{Perrotta2019} for detailed discussions).

While the derived outflow is extreme, it provides the best explanation for the relatively flat spectrum between rest-frame 4900--5000\,\AA.
If no outflow component is included in the spectral model, large residuals are present around \oiiib, with this wavelength region significantly under-fit. 
There may be some contribution in this region from other transitions such as \ions{He}{i} \citep[e.g.][]{Uebler2023} that cannot be excluded because of the low spectral resolution. The \ions{Fe}{ii} template may also not be an ideal model for this quasar. However, neither explanation is likely to account for the significant flux around \oiiib, and a broad outflow is the most likely scenario. 
This \hb--\Oiii\ spectral shape is very similar to that of the $z\simeq2$ extremely red quasar J102541.78+245424.2, with relatively constant flux between rest-frame 4900--5000\AA, that was measured to have a similar extreme outflow with \oiiia\ FWHM of $5751\pm195$\,km\,s$^{-1}$ \citep{Perrotta2019}, further supporting this outflow scenario.

Following \citet{Carniani2015}, we calculate the outflow masses from the \oiiia\ and \hb\ luminosities, assuming a density of the gas in the ionised outflow $\langle n_e\rangle = 500\,\rm{cm}^{-3}$.
From the \oiiia\ FWHM maps in Figure \ref{fig:OIIImaps}, this broad emission is not present after PSF subtraction, implying that it is spatially unresolved. We therefore assume an outflow radius $R_{\rm{out}}=1\pm0.1$ kpc, approximately the spatial resolution of JWST, as in \citet{Zamora2024}.
The estimated ionised outflow rate $\dot{M}_{\rm{out}} = M_{\rm{out}} v_{\rm{out}}/R_{\rm{out}}$
and the kinetic power of the outflow $\dot{P}_{\rm{K,out}} = \frac{1}{2}\dot{M}_{\rm{out}} v^2_{\rm{out}}$
are listed in Table~\ref{tab:Outflows}.
\hb\ gives a larger outflow mass, rate, and kinetic power than \oiiia. This is a known consequence of the ionisation structure of AGN NLR clouds, with \oiiia\ tracing only a small portion of the mass of each cloud (e.g. \citealt{Perna2015}, as also seen in the high-$z$ quasars of \citealt{Marshall2023}).

The kinetic power of the outflow is $33\substack{+5\\-25}$\% (from \hb) and $8\pm1$\% (from \oiiia) of the bolometric luminosity of the quasar.
These measurements suggest that \NDWFS\ has 
a very powerful quasar-driven ionised-gas outflow, 
potentially one of the most extreme 
in the early Universe. This outflow would likely quench the star formation in the host galaxy.
However, these measurements rely on the model fit to the low-resolution spectral data, in which the various emission line components are highly blended. 
More accurately constraining these outflow properties and confirming its extreme nature requires higher spectral resolution data, and these results are only tentative evidence of such an extreme outflow.

\section{Extended emission structures}
\label{sec:GalaxyResults}
\subsection{Spectroscopic confirmation of two companion galaxies and discovery of a gas bridge}

\begin{figure}
\begin{center}
\includegraphics[scale=0.86]
{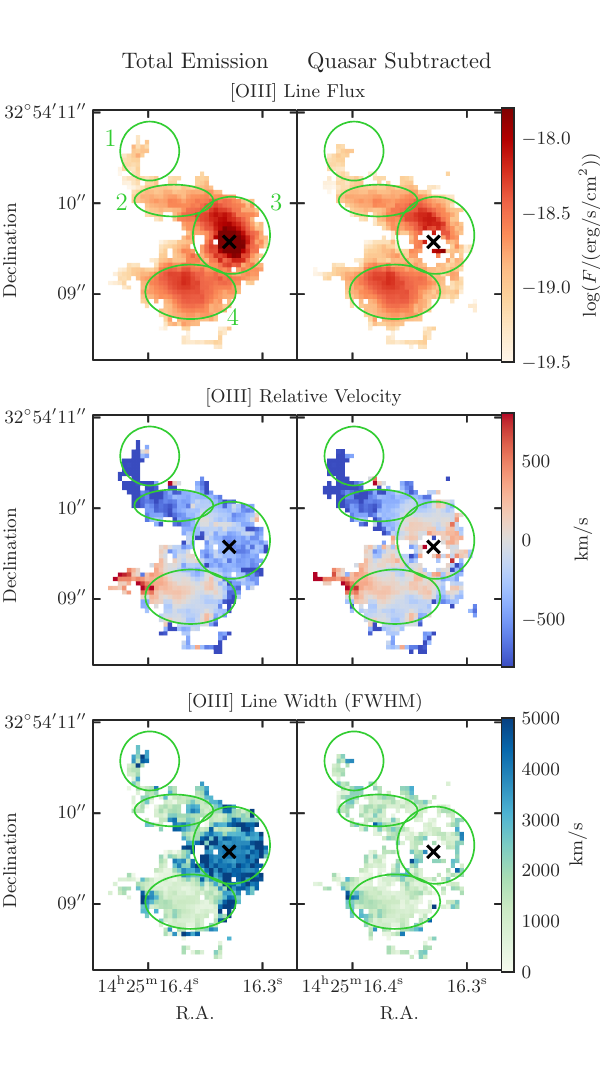}
\caption{Maps of the \oiiia\ emission line regions surrounding \NDWFS. From top to bottom, the panels show the line flux, velocity, and line width. 
The left column shows the data cubes containing both the quasar and extended emission, while the right column shows the cubes after the quasar has been subtracted. 
The location of the quasar is marked as a black cross, and the emission line region apertures from Figure \ref{fig:HST} are included for reference.
The line velocities and widths are measured from Gaussian fits to the lines in each spaxel; the velocity is measured as the line centre relative to the quasar host redshift of $z=5.901$, and the width is the FWHM corrected for the instrumental resolution of 1966\,km\,s$^{-1}$. 
}
\label{fig:OIIImaps}
\end{center}
\end{figure}

Figure \ref{fig:OIIImaps} shows the \oiiia\ flux and velocity maps for the system, both before and after subtracting the quasar's contribution.
The brightest region of \oiiia\ is emitted from the quasar, following the PSF shape, with line $\rm{FWHM}\gtrsim4000$\,km\,s$^{-1}$.
There are also extended regions of \oiiia\ emission in the south and north-east that are extended beyond the PSF shape and are clearly seen even before quasar subtraction. After quasar subtraction, some \oiiia\ emission is also seen from the quasar host galaxy, although this is difficult to detect relative to the bright quasar emission. Despite the quasar subtraction resulting in a core of artefacts in the close vicinity of the quasar, the subtracted image gives a much clearer picture of the kinematics of the extended emission line regions.

Figure \ref{fig:HST} shows the \oiiia\ flux distribution after quasar subtraction, alongside the integrated 0.82--5.2\micron\ flux map and the quasar-subtracted HST/WFC3 F125W image \citep{Marshall2019c}. 
The north-east corner of the data cube exhibits faint \oiiia\ emission coincident with a bright region of continuum emission.  
From the WFC3 F125W and F160W imaging of this north-eastern source, \citet{Marshall2019c} measured $m_J=24.6\pm0.1$\,mag and $m_J-m_H=0.4\pm0.1$\,mag. This corresponds to a UV absolute magnitude $M_{1500}=-21.8\pm0.2$\,mag and UV slope $\beta=-0.4\pm0.8$, assuming the source is at the same redshift as the quasar.
\citet{Marshall2019c} fit the source with a S\'ersic profile, finding a best-fitting index $n=3.6\pm0.7$, radius $R_e=2.6\pm0.4$ kpc, and axis ratio $b/a=0.81\pm0.21$ at a projected distance of ${\sim}1\farcs4$ or $8.4\pm0.1$\,kpc from the quasar.
Based on the UV slope and magnitude, \citet{Marshall2019c} concluded that this source was likely a high-$z$ galaxy, not a foreground interloper.
Our IFU spectra confirm that this source is indeed a companion galaxy at the same redshift as the quasar.

Figure \ref{fig:HST} shows a region of \oiiia\ emission connecting the quasar host galaxy and the north-east companion. 
This `region~2' has much larger \oiiia\ flux than region~1, yet no continuum emission from this region was detected in the F125W or F160W HST imaging. The lack of continuum detection suggests that region~2 is most likely a connecting bridge of hot, ionised gas between the host galaxy and the north-eastern companion.

To the south-east of the quasar, \citet{Marshall2019c} discovered a bright galaxy in the F125W and F160W imaging, with $m_J=24.4\pm0.1$\,mag and $m_J-m_H=0.1\pm0.1$\,mag. These correspond to 
UV magnitude 
$M_{1500}=-22.3\pm0.2$\,mag and 
UV slope 
$\beta=-1.6\pm0.6$.
The HST imaging for this companion is best fit with a S\'ersic profile with 
index of 
$n=0.5\pm0.1$, 
radius 
$R_e=2.7\pm0.1$ kpc, and 
axis ratio 
$b/a=0.94\pm0.07$ at a projected distance of ${\sim}0\farcs6$ or $3.4\pm0.2$ kpc from the quasar.
All of these results assume that the source is at the same redshift as the quasar, but
\citeauthor{Marshall2019c} cautioned that the southern source might be too bright to be a galaxy at the same redshift as the quasar.
Our new IFU data give a clear detection of \oiiia\ at the same south-east location (`region~4'), confirming that this is indeed a very bright companion galaxy of the quasar.

Overall, we were able to spectroscopically confirm the presence of two companion galaxies around the quasar and discover a bridge of gas connecting the northern companion to the quasar host. We now study the physical properties of these regions.

\subsection{Gas properties of the emission-line regions}
\label{sec:FluxMeasurements}

\begin{figure*}
\begin{center}
\includegraphics[scale=0.86]{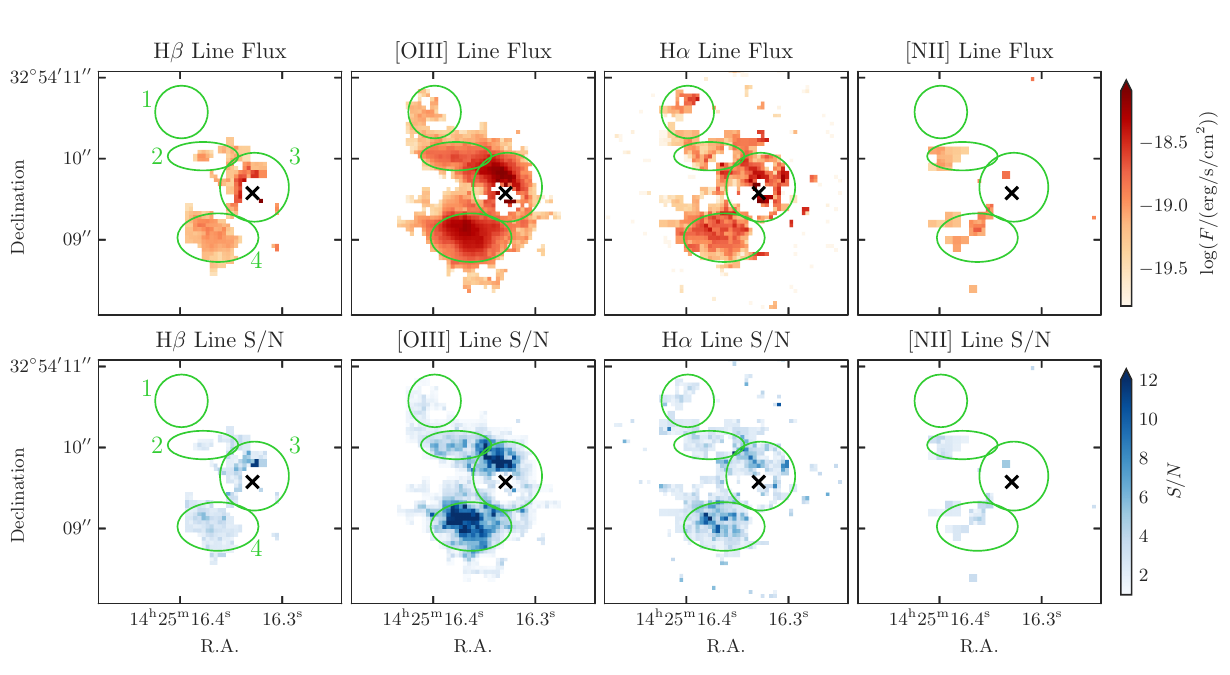}
\caption{Maps of the emission-line regions surrounding \NDWFS\ after the quasar emission has been subtracted. From left to right, the panels show the \hb, \oiiia, \ha, and \niia\ lines, with the upper panels showing the flux in each spaxel and the lower panels showing their $S/N$, according to the colour bars at the right of each row. The \oiiia\ and \niia\ fluxes are from fits with the doublet flux ratios constrained (Table~\ref{tab:HostFlux} note). The location of the quasar is marked as a black cross, and the emission-line region apertures from Figure~\ref{fig:HST} are marked in green.}
\label{fig:FluxMaps}
\end{center}
\end{figure*}

Figure \ref{fig:FluxMaps} shows the flux and $S/N$ maps for \oiiia, \hb, \ha, and \niia, after quasar subtraction. 
For these maps, we usec a hybrid resolution in order to see fainter flux regions. As for Figure~\ref{fig:OIIImaps}, the maps include spaxels from the 0\farcs05 pixel-scale cube where the emission line $S/N>3$ and any adjacent spaxels with $S/N>1.5$. However, we also binned this data cube onto a 0\farcs1 pixel scale by summing the spectra within each set of 2$\times$2 spaxels and refitting the resulting spectra as a series of Gaussians as for the 0\farcs05 cube (Section \ref{sec:MapFitting}).
We created a second mask including spaxels that have $S/N>5$ and any adjacent spaxels with $S/N>2$ in the 0\farcs1 cube. For spaxels that were not selected in the 0\farcs05 cube mask, we applied the mask from the 0\farcs1 cube for the corresponding spaxel.
This reveals regions that have too low $S/N$ to be detected in the high-resolution cube: \hb\ and \ha\ in the outskirts of the emission line regions, and particularly the faint \niia\ line in regions~2 and~4.

The \oiiia\ emission has the largest $S/N$ and therefore is the clearest tracer of the emission structures. \ha\ generally follows the \oiiia\ spatial distribution, although for region 1, the \ha\ emission is brightest in the north while the \oiiia\ emission is brightest in the south-east. This indicates that these lines may be emitted from different gas structures within the northern companion, but higher $S/N$ would be required to confirm this.
The $S/N$ for \hb\ and \niia\ are too low for these to be detected across the whole system on a spaxel-by-spaxel basis. 
\hb\ is primarily detected around the host galaxy and the southern companion, with some faint emission in the connecting gas bridge.
\hb\ is not seen in the west of the host galaxy, where \oiiia\ and \ha\ are detected; this is likely due to a low $S/N$ of the fainter \hb\ line as well as difficulty in extracting the line after quasar subtraction.
As discussed below, \nii\ emission is difficult to resolve from the \ha\ line. However, this fitting algorithm, which constrains each emission line to have the same velocity width, results in the tentative detection of \niia\ throughout the gas bridge and within the southern companion.

\begin{figure*}
\begin{center}

\includegraphics[scale=0.86]{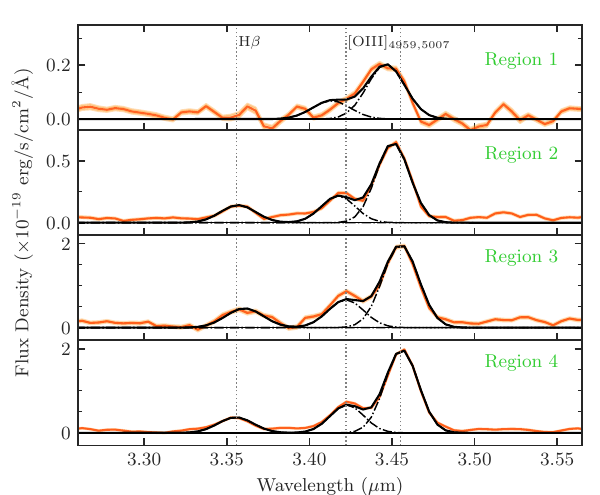}
\hspace{-0.4cm}
\includegraphics[scale=0.86]{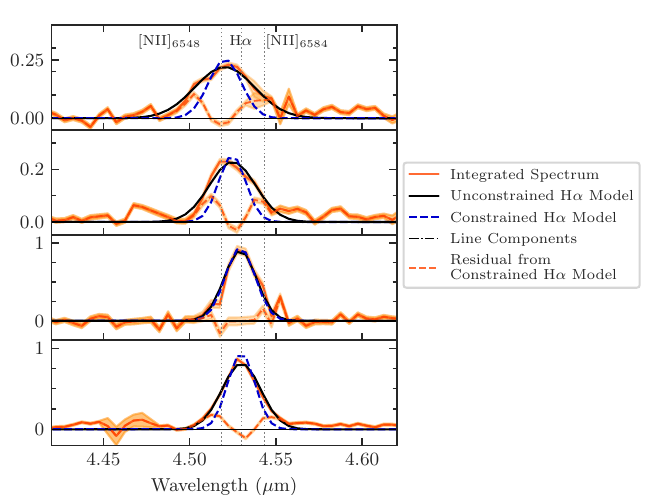}
\caption{Integrated spectra for the emission-line regions surrounding \NDWFS\ (orange solid), showing the \hb--\Oiii\ wavelength range (left panel) and the \ha--\Nii\ wavelength range (right panel). The blue dashed line shows the best model assuming that the physical line width of \ha\ is fixed to that of \oiiia:\  the `constrained \ha\ model'. The black solid line shows the best model allowing the \ha\ line width to vary:\ the `unconstrained \ha\ model.'
The dashed orange line shows the residual around \ha\ after subtracting the constrained \ha\ model.
The orange shaded region shows the $\pm1\sigma$ uncertainty level of the spectra. 
The individual Gaussian model components of \hb\ and \Oiii\ are shown as dashed lines. The \oiiia/\oiiib\ ratio is constrained to 2.98.
The dotted vertical lines mark the emission line locations at the quasar host redshift, $z=5.901$.
}
\label{fig:Regions}
\end{center}
\end{figure*}

To obtain a single, higher $S/N$ spectrum of each region, we integrate this spectral cube within each of the Figure~\ref{fig:HST} region apertures. 
Figure~\ref{fig:Regions} shows these spectra across the wavelengths covering the relevant lines.
We fit each integrated spectrum with a single Gaussian for each of the emission lines \oiii, \hb, and \ha; we do not fit the \nii\ lines.
The \oiii\ and \hb\ lines are tied to have the same central velocity and physical velocity width, and the \oiii\ doublet flux ratio is fixed as in Section~\ref{sec:MapFitting}. 
For regions~2 and~3, we allow the \hb\ central velocity to vary, as this provides a significantly improved fit to the data.
For \ha, we fit two separate models in order to determine whether there is any significant detection of the blended \nii\ lines. The first model fixes the \ha\ physical line centre and width to be the same as \oiiia, corrected for instrumental resolution:\ the `constrained \ha' model. This is the more physically realistic model, as lines emitted from the same gas would have the same physical velocity structure. The second model allows for the \ha\ physical line width to vary:\ the `unconstrained \ha' model. This model represents the case that all of the flux in the blended emission line is from \ha, with no contribution from \nii.
Table~\ref{tab:HostFlux} gives the resulting line fluxes and ratios, the Gaussian central velocity $v_r$, and the Gaussian velocity width $v_\sigma$, corrected for instrumental resolution. The \ha\ flux and velocity width from the unconstrained \ha\ model, $F_{\rm H\alpha, max}$ and $v_{\sigma, \rm H\alpha, max}$, are the maximal \ha\ fluxes and velocity widths assuming no contribution from \nii. The \ha\ flux from the constrained \ha\ model is $F_{\rm H\alpha, min}$.

\begin{table*}
\caption{Fluxes and velocities for the \hb, \oiiia, \ha, and \niia\ galaxy line components for each emission line region as defined in Figure \ref{fig:HST}, measured from the integrated spectra shown in Figure \ref{fig:Regions}. 
}
\begin{center}
\begin{footnotesize}
\begin{tabular}{lrrrr}
\hline
Quantity & Region 1 & Region 2 & Region 3 & Region 4 \\
\hline
$F_{\rm H\beta}$ & $<$1.4 ($3\sigma$) & 3.7$\pm$0.3 [11] & 12.6$\pm$2.0 [6] & 9.6$\pm$0.4 [22]\\
$F_{\rm [O\textsc{iii}]}$ & 5.8$\pm$0.4 [15] & 16.3$\pm$0.4 [43] & 53.2$\pm$1.7 [31] & 49.9$\pm$0.5 [95]\\
$F_{\rm H\alpha, min}$ & 6.8$\pm$0.6 [11] & 5.2$\pm$0.3 [16] & 24.1$\pm$2.1 [12] & 19.6$\pm$0.5 [42]\\
$F_{\rm H\alpha, max}$ & 12.4$\pm$1.4 [9] & 10.1$\pm$0.8 [13] & 27.9$\pm$2.5 [11] & 30.1$\pm$0.9 [34]\\
$F_{\rm [N\textsc{ii}]}$ & $<$1.2 ($3\sigma$) & $\leq$0.7$\pm$0.2 [4] & $<$2.3 ($3\sigma$) & $\leq$2.2$\pm$0.6 [4]\\
$F_{\rm H\alpha}/F_{\rm H\beta}$ & $>$3.5 ($3\sigma$) & $\leq$2.7$\pm$0.3 & $\leq$2.2$\pm$0.4 & $\leq$3.1$\pm$0.2\\
$F_{\rm [O\textsc{iii}]}/F_{\rm H\beta}$ & $>$3.2 (3$\sigma$) & 4.4$\pm$0.4 & 4.2$\pm$0.7 & 5.2$\pm$0.2\\
$F_{\rm [N\textsc{ii}]}/F_{\rm H\alpha}$ & $<0.18$ & $\leq$0.14 & $<$0.10 & $\leq$0.11\\
$v_r$ & $-762$$\pm$217 & $-399$$\pm$217 & $0$$\pm$217 & $57$$\pm$217\\
$v_{r, \rm H\beta}$ & - & $190$$\pm$217 & $524$$\pm$217 & $51$$\pm$217\\
$v_\sigma$  & $535\substack{ +405\\-535 }$ & $254\substack{ +528\\-254 }$ & $416\substack{ +447\\-416 }$ & $255\substack{ +527\\-255 }$\\[1ex]
$v_{\sigma, \rm H\alpha, max}$ & $866\substack{ +204\\-228 }$ & $730\substack{ +213\\-252 }$ & $800\substack{ +208\\-238 }$ & $731\substack{ +213\\-252 }$\\[1ex]

\hline
\end{tabular}
\end{footnotesize}
\end{center}
\tablefoot{Fluxes are quoted in units of $10^{-18}$\,erg\,s$^{-1}$\,cm$^{-2}$, and values in square brackets show the integrated line $S/N$\null. Velocities are quoted in units of km~s$^{-1}$. The \oiii\ doublet is constrained to have a flux ratio of \oiiia/\oiiib\ $=2.98$; the tabulated fluxes are for \oiiia.
The constrained \ha\ model fluxes are $F_{\rm H\alpha, min}$, with this model constrained such that \ha\ has the same velocity dispersion as \oiiia, $v_{\sigma}$.
The unconstrained \ha\ model fluxes and velocity dispersions are listed as $F_{\rm H\alpha, max}$ and $v_{\sigma, \rm H\alpha, max}$, which assume no contribution from \nii\ to the blended line. 
For the undetected \hb\ flux in region 1 and the \nii\ fluxes in regions 1 and 3, the 3$\sigma$ limits are from the rescaled ERR array accounting for the quasar subtraction uncertainty (Section~\ref{sec:quasarSubtraction}).
For regions 2 and 4, where there is tentative evidence of the presence of \nii, the table gives the \niia\ flux as the measured residual after subtracting the constrained \ha\ model, but the result is treated as an upper limit.
Our quoted $F_{\Nii}/F_{\rm H\alpha}$ ratios use these \niia\ upper limits with $F_{\rm H\alpha, min}$ to give an upper limit on the ratio.
The central velocities $v_r$ are defined relative to the host central velocity at $z=5.901$.
Uncertainties on the velocity and velocity dispersion assume a measurement uncertainty of 0.5 wavelength elements, $\pm25$\,\AA, which corresponds to 217\,km\,s$^{-1}$ at \oiiia. 
The velocity dispersions are the physical velocity widths, corrected for instrumental resolution (where FWHM 
$v_{\rm{inst}}=1966\pm178$\,km\,s$^{-1}$ for \oiiia\ from Appendix \ref{sec:LSF}; these uncertainties are incorporated into the velocity-dispersion uncertainty).
}
\label{tab:HostFlux}
\end{table*}

Region~1, which has the faintest emission lines, has significant detections of \oiiia\ and \ha\ but no detection of \hb. All other regions have significant \ha, \hb, and \oiii\ detections.
It is difficult to determine whether \nii\ emission is present because these lines are blended with \ha\ at this spectral resolution. For regions~1 and~3, the constrained \ha\ model has residuals with $S/N<3$ around \nii, implying no detection of \Nii. For these regions, we quote the \niia\ upper limit as $3\sigma$ from the noise array.
For regions~2 and~4, the residuals of the constrained \ha\ model at the locations of \nii\ have  $S/N\simeq4$--5. However, if the residuals are purely from \Nii, we would expect the \niia\ flux to be 3.05 times the \niib\ flux \citep{Dojcinovic2023}, yet we see no such asymmetry in the residuals. 
We therefore have only tentative evidence of the presence of \Nii\ in regions 2 and 4, and treat the measured residual fluxes as upper limits.  These may be  \Nii\ detections, although noise and an imperfect \ha\ model are likely contributing significantly to this measured residual.

\subsubsection{Photoionisation mechanisms and dust attenuation}
\label{sec:BPT}

\begin{figure}[tbh]
\begin{center}
\includegraphics[scale=0.86]{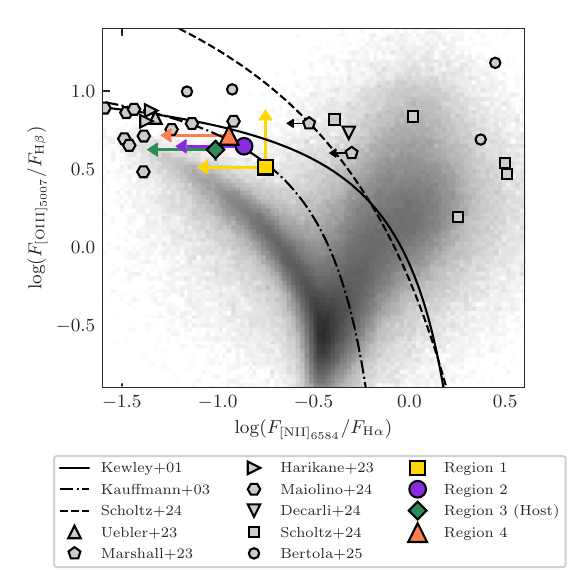}
\caption{BPT diagnostic diagram \citep{Baldwin1981} showing the narrow emission line ratios $F_{\rm{[N\textsc{ii}]{\lambda6584}}}/F_{\rm{H\alpha}}$ and $F_{\rm{[O\textsc{iii}]{\lambda5007}}}/F_{\rm{H\beta}}$ for the four regions (coloured data). 
Our quoted $F_{\rm [N\textsc{ii}]}/F_{\rm H\alpha}$ ratios use the \niia\ upper limits with $F_{\rm H\alpha, min}$, giving an upper limit on the ratio.
The \citet{Kewley2001}, \citet{Kauffmann2003}, and \citet{Scholtz2023} demarcation curves distinguish the regions where photoionisation is dominated by galaxies (lower left) and AGNs (upper right), as determined for low-$z$ objects. The underlying grey histogram shows local sources from the SDSS \citep{Abazajian2009}.
The grey points show a sample of $3.5<z<10$ AGN and quasar host galaxies measured with JWST \citep[][using the combined narrow and outflow fluxes for sources that were kinematically decoupled into two components to compare to our single-component line fits]{Uebler2023,Marshall2023,Harikane2023,Maiolino2023,Decarli2024,Scholtz2023,Bertola2025}.
}
\label{fig:BPT}
\end{center}
\end{figure}

The available emission line ratios offer several diagnostics to understand the nature of the various regions.
The classical BPT diagram \citep{Baldwin1981} 
can determine the dominant cause of gas photoionisation at low-$z$. 
Figure~\ref{fig:BPT} presents the BPT diagram for the regions.
All four regions have high $F_{\rm [O\textsc{iii}]}/F_{\rm H\beta}$ and low $F_{\rm [N\textsc{ii}]}/F_{\rm H\alpha}$, placing them in the area of the  diagram mostly occupied by star-forming galaxies at low-$z$. However, at high-$z$, where metallicities are low, AGNs can also lie in this region of the diagram \citep[e.g.][]{Groves2006,Nakajima2022,Dors2024,Cleri2025}.
Figure~\ref{fig:BPT} includes some examples of observed high-$z$ AGNs and quasars, showing that the four regions are consistent with being photoionised by high-$z$ AGN\null.
There is no evidence of a point-source emitter suggestive of a second type~1 AGN in either companion galaxy. If these regions are AGN-photoionised, this is most likely caused by the quasar itself, although an obscured AGN in one or both companions cannot be ruled out.
Other diagnostic lines such as [\ions{S}{ii}]\,$\lambda\lambda$6716, 6731, [\ions{O}{i}]\,$\lambda$6300,  [\ions{O}{ii}]\,$\lambda$3727,  [\ions{O}{iii}]\,$\lambda$4363,  \ions{He}{ii}\,$\lambda$4686, and/or [\ions{Ne}{v}]$\,\lambda\lambda$3346, 3426 \citep[e.g.][]{Veilleux1987,Shirazi2012,Mignoli2013,Mazzolari2024}
are needed to determine the photoionisation mechanism, but these lines are not detected in our spectra.

The nebular dust attenuation can be measured from the Balmer decrement, the observed flux ratio of narrow \ha\ to \hb.
The theoretical non-extincted ratio is $F_{\rm{H}\alpha}/F_{\rm{H}\beta} \simeq 2.86$, assuming a temperature of 
$T = 10^4$ K and an electron density $n_{\rm e} = 10^2
~\rm{cm}^{-3}$
for Case~B recombination \citep{Hummer1987}. 
In regions where photoionisation has a significant contribution from an AGN, additional collisional excitation of \ha\ results in a \ha/\hb\ ratio larger than 2.86 \citep{Ferland1983,Halpern1983,Osterbrock1989}, 

and so we expect an intrinsic flux ratio of $\geq$2.86 for our four regions.
Table~\ref{tab:HostFlux} lists the flux ratios, taking an upper limit based on $F_{\rm H\alpha, max}$ from the unconstrained \ha\ model. For regions~2 and~4, the \ha/\hb\ flux ratios are consistent with 2.86 within $1\sigma$. For region~3, the \ha/\hb\ flux ratio is less than 2.86 albeit consistent within $2\sigma$. 
Due to the blending of \ha\ and \Nii, the complexity of quasar subtraction, and the resulting low S/N of our emission lines, we do not have precise measurements of the \ha/\hb\ flux ratio. However, we see no significant evidence of the presence of dust in these three regions from their Balmer decrements, as also suggested by their blue spectra (Figure~\ref{fig:pPXF}).

In contrast to the other regions, region~1 has no \hb\ detection, and $F_{\rm{H}\alpha}/F_{\rm{H}\beta}>3.5$ (3$\sigma$). This is based on $F_{\rm H\alpha, min}$ as $F_{\rm H\alpha, max}$ would give a larger lower limit.
Following \citet{Dominguez2013}, the nebular colour excess  $E(B-V) = 1.97 \log_{10}[ (F_{\rm{H}\alpha}/F_{\rm{H}\beta}) / 2.86]>0.17$\,mag, and 
the \citet{Calzetti2000} reddening law with $R_V=4.05$ gives
$A_{\rm{H}\alpha}>0.52$~mag. Therefore, there is likely some level of dust attenuation in the northern companion galaxy, although we cannot precisely correct for this given our lack of detection of \hb.

Understanding the photoionisation mechanisms and dust properties of this system requires deeper and higher-resolution spectra. 
Higher $S/N$ spectra would capture the \hb\ line of region~1 and more precisely measure the line fluxes and ratios.
Higher spectral resolution data would de-blend the \Nii\ and \ha\ lines, which would give an uncontaminated measurement of the \ha\ flux and the \Nii/\ha\ flux ratios.

\begin{table*}
\caption{Physical properties calculated for the emission line regions: SFR from the constrained (SFR$_{\rm H\alpha, min}$) and unconstrained (SFR$_{\rm H\alpha, max}$) \ha\ models, mass-to-light ratio, $M_\ast/L$, in the HST/WFC3 F160W filter, stellar age, $t$, metal abundance, $[M/H]$, stellar dust attenuation, $A_V$, and stellar mass, $M_\ast$.
}
\begin{center}
\begin{footnotesize}
\begin{tabular}{lrrrrrrrr}
\hline 
Region 
& SFR$_{\rm H\alpha, min}$
& SFR$_{\rm H\alpha, max}$
 & $M_\ast/L$
 & $t$
 & $[M/H]$ 
 & $A_V$
 & $M_\ast$\\
& (\Msol\,yr$^{-1}$) & (\Msol\,yr$^{-1}$) & ($10^{-4}$ $M_\odot/L_\odot$) & (Myr) & & (mag) & ($10^{10}$\,\Msol)\\
\hline
1 & 14.7$\pm$1.3 & 26.7$\pm$3.0 & $14\substack{+3\\-1}$ & $65\substack{+9\\-4}$ & ${>}{-0.1}$ & $0.41\substack{+0.04\\-0.05}$ & $36\substack{+6\\-3}$
\\
2 (2*) & 11.2$\pm$0.7 & 21.8$\pm$1.7  & $1.6\substack{+2.0\\-0.7}$ & $8\substack{+10\\-4}$ & -- & -- & -- 
\\
3  & 52.1$\pm$4.6 & 60.4$\pm$5.1  & -- & --& --& -- & --
\\
4 (4*) & 42.4$\pm$1.0 & 65.1$\pm$1.9  & $0.7\substack{+0.2\\-0.1}$ & $6.7\substack{+1.8\\-1.8}$ & ${<}-0.8$ & $1.2\substack{+0.1\\-0.1}$ & $1.9\substack{+0.4\\-0.4}$
\\
\hline
\end{tabular} 
\end{footnotesize}
\end{center}
\tablefoot{The SFRs are based on the \ha\ narrow-line fluxes listed in Table \ref{tab:HostFlux} following the equation in Section \ref{sec:SFRs}.
The true SFRs are likely to be smaller than the calculated values because the quasar contributes to the photoionisation.
The stellar population properties $t$, $[M/H]$, and $A_V$ are calculated from the \textsc{pPXF} SPS modelling (Section~\ref{sec:pPXF}). These models use the smaller regions~2* and~4* to avoid contamination by the quasar.
These SPS quantities are light-weighted values.
The stellar mass $M_\ast$ is calculated from the SPS mass-to-light-ratio and the HST/WFC3 F160W $H$-band magnitude of the two companion galaxies.
Region~2 was undetected in the $H$-band images, prohibiting a stellar mass estimate.
Region~3 is dominated by the quasar emission, and no SPS fitting is possible.
}
\label{tab:PhysicalProperties}
\end{table*}

\subsubsection{Star formation rates}
\label{sec:SFRs}
If the \ha\ emission is excited by ionising photons from hot stars, the \ha\ flux measures the star-formation rate (SFR)\null. 
With no dust attenuation, a solar abundance, and a \citet{Kroupa2001} IMF, the SFR can be estimated as \citep{Kennicutt2012}:\ 
\begin{equation}
\textrm{SFR} = 5.37\times 10^{-42} \frac{L_{\rm{H}\alpha}}{(\textrm{erg\,s$^{-1}$})}\,\rm{M}_\odot\,\rm{yr}^{-1}
\label{eq:Kenn}
.\end{equation}
Table~\ref{tab:PhysicalProperties} lists the implied SFRs for each region,
from both the constrained and unconstrained \ha\ models.
If an AGN contributes to the photoionisation of any region, that region's SFR will be overestimated.

For the host galaxy (region~3), the measured \ha\ SFR is $\sim50$--$60\,M_\odot$\,yr$^{-1}$.
This is far less than the SFR estimate from the FIR luminosity \citep{Wang2010} of 1160$\,M_\odot$\,yr$^{-1}$. On its face, this comparison suggests that most of the star formation in the quasar host is heavily obscured, implying that there must be significant dust. While the Balmer decrement gives no evidence of any significant dust attenuation of \ha, the spectral measurements are complicated by the quasar subtraction process, low $S/N$, and blending of \Nii\ and \ha, and therefore dust obscuration cannot be ruled out. Many effects make the comparison between FIR and \ha\ SFRs difficult: the FIR and \ha\ trace star formation on different timescales, with the FIR sensitive to star formation to $\gtrsim$100\,Myr ago, while the \ha\ captures stars formed over only the past ${\sim}$10\,Myr \citep{Kennicutt2012}; the FIR observations have poor spatial resolution and include the flux from all four regions; and there is potential for significant contamination by the quasar for both measurements \citep[e.g.][]{Tsukui2023}.

\NDWFS\ was originally targeted in the HST programme due to its bright FIR luminosity and relatively faint UV luminosity (relative to comparable high-$z$ quasars), with $F_{250\rm{GHz}}/F_{1\micron}\simeq100$ \citep{Marshall2019c}. 
A high FIR luminosity implies a large SFR and, thus, an intrinsically bright host, which combined with the lower UV luminosity may result in a less extreme quasar-to-host contrast ratio and thus improved detectability of the host via quasar subtraction.
This selection is consistent with significant ongoing, but obscured, star formation. With such a FIR excess, it is unsurprising that this luminous infrared galaxy (LIRG)-like quasar host is undergoing galaxy mergers, as LIRGs are commonly found to have merger-induced star formation and AGN activity \citep[e.g.][]{Veilleux2002,Younger2009}.

The south-east companion (region 4) has significant \ha\ emission, and unless this is powered by the AGN, this galaxy must have moderate ongoing star formation of $\sim$40--65\,M$_\odot$\,yr$^{-1}$.
The north-east neighbour (region~1) has weaker observed \ha\ flux with corresponding measured SFR of $\sim$15--30\,M$_\odot$\,yr$^{-1}$. This region has measured dust attenuation  $A_{\rm{H}\alpha}>0.52$~mag making the dust-corrected SFR  ${>}24\,M_\odot$\,yr$^{-1}$ for the constrained \ha\ model and ${>}29\,M_\odot$\,yr$^{-1}$ for the unconstrained \ha\ model. This region could be undergoing significant obscured star formation.
The bridge (region 2) has the weakest measured \ha\ flux, which could be produced by star formation of $\sim11$--22\,M$_\odot$\,yr$^{-1}$.

\subsubsection{Kinematics}
As the spectral resolution of these observations is only $R\sim100$, all measured velocities and velocity dispersions have significant uncertainty, as well as there being significant uncertainty in the measured LSF\null. Nevertheless, the relative velocity trends shown in the \oiiia\ maps (Figure~\ref{fig:OIIImaps}) should be accurate. The northern companion is blueshifted relative to the quasar host, there is a velocity gradient across the connecting bridge extending to the velocity of the quasar host, and the southern companion has a redshifted velocity relative to the quasar. There is significant spatially unresolved broad \oiiia\ emission from the quasar, likely from a strong outflow, while the remainder of the emission structures exhibit much lower velocity dispersions.
However, the specific velocity quantities are uncertain, as reflected by the quoted uncertainties. Indeed the velocity dispersions for each region from Table \ref{tab:HostFlux} are all consistent with zero within $1\sigma$. Higher spectral resolution data of this system, particularly with the NIRSpec IFU $R\sim2700$ grating, would allow for a more quantitative study of the velocities within this system \citep[see e.g.][]{Arribas2024,Fujimoto2024,Jones2024,Marconcini2024,Parlanti2024a,Parlanti2024,DelPino2024,Uebler2024b}.
Given the uncertainties, we generally avoid placing significance on the measured velocities in this work and, instead, we focus on the line fluxes and associated physical properties.

\subsection{Stellar properties of the emission-line regions}
\label{sec:pPXF}

\begin{figure}[tbh]
\begin{center}
\includegraphics[scale=0.86]{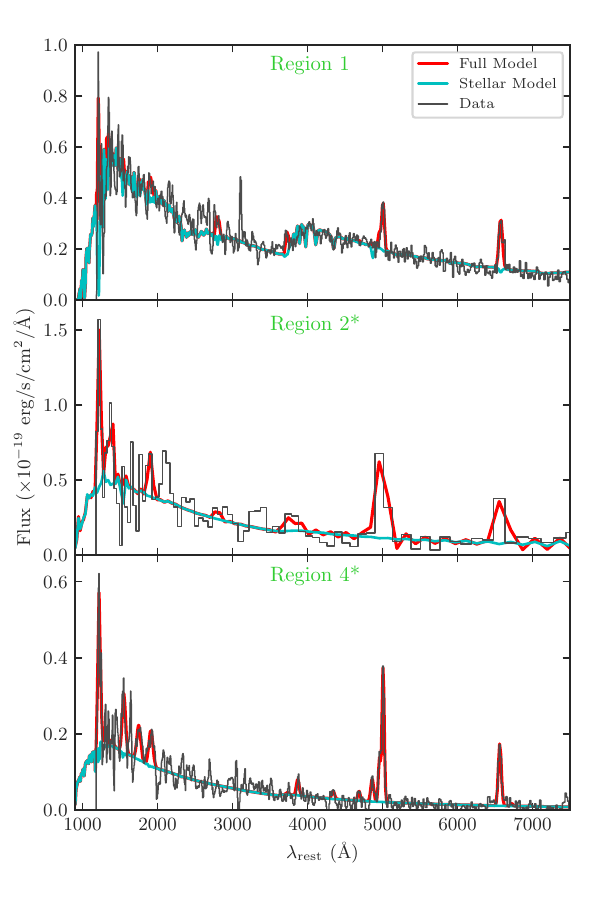}
\caption{Spectra for regions 1, 2*, and 4* (black), integrated over the apertures shown in Figure \ref{fig:PSFs}. 
The red curve shows the full gas and stellar model as fit by \textsc{pPXF} for each region with the stellar component shown in blue. 
These spectra are taken from the original data cube, which includes the continuum emission and the quasar flux. The apertures for regions 2* and 4* are smaller than regions 2 and 4 to exclude flux from the quasar and thus ensure an accurate measurement of the stellar population.
The spectrum for region 2* is rebinned on a $10\times$ coarser wavelength grid to obtain a higher $S/N$ because the continuum has $S/N\lesssim5$ on the original grid.
}
\label{fig:pPXF}
\end{center}
\end{figure}

Figure \ref{fig:pPXF} shows the integrated spectra for regions~1, 2*, and 4* (the latter two being the smaller subsets of regions~2 and~4 shown in Figure~\ref{fig:PSFs}). These data, used for fitting the continuum emission with \textsc{pPXF}, have two main differences from the spectra in Figure~\ref{fig:Regions}. Firstly, these spectra 
are taken from the original cube containing the continuum emission.
Secondly, regions 2* and 4* are smaller than regions~2 and~4, defined to exclude most of the quasar emission and thus give an un-contaminated, accurate spectral shape for the companion sources. This enables SPS modelling, although the smaller area does not contain all of the flux of each source, resulting in a lower $S/N$.
Region~3 cannot be studied in this way because the quasar contaminates the entire source.

We use \textsc{pPXF} to fit stellar population models to these spectra as described in Section \ref{sec:SPSmodelling}, with Table~\ref{tab:PhysicalProperties} listing the output ages and metallicities of the stellar populations.
All of the companion regions show clear continuum emission across the rest-frame UV, with region~1 also showing significant emission at rest-frame wavelengths $\ga$3800~\AA\null.
The northern companion region~1
shows a clear Balmer break from the older stellar population. The metallicity is poorly constrained but relatively high.
For the connecting bridge, region~2*, the low $S/N$ and spectral resolution mean \textsc{pPXF} cannot constrain the metal abundance or stellar dust attenuation, but the blue UV slope and lack of a Balmer break suggest a young stellar population. 
For the southern companion, region 4*, \textsc{pPXF} measures a similarly young stellar age,
modest but non-zero stellar dust attenuation, and low metal abundance.

From these stellar population models and the resulting mass-to-light ratio estimates, we were able to estimate the stellar masses for the emission regions.
We could not determine the total flux of these regions from the IFU spectra, due to contamination by the quasar emission.
We therefore used HST/WFC3 photometry with the quasar flux subtracted \citep{Marshall2019c} to give accurate $H$-band fluxes of the northern and southern companions: $m_H=24.2$\,mag for region 1, and $m_H=24.3$\,mag for region 4.
Table~\ref{tab:PhysicalProperties} gives the resulting stellar masses and the estimated mass-to-light ratios from the SPS fitting from which they are calculated.
The connecting gas bridge was undetected in the WFC3 images, and so no stellar mass estimate is possible in this manner.

The northern companion has an older, massive stellar population that is relatively metal-rich, 
while the southern companion is younger, less massive, and metal-poor.
The stellar mass of the northern companion, $(3.6\substack{+0.6\\-0.3})\times10^{11}\,M_\odot$, is equivalent to the most massive galaxies observed at these redshifts \citep{Xiao2024}. Its SFR of 15--30$\,M_\odot$\,yr$^{-1}$ puts the northern companion $\sim$1.1\,dex below the star-forming main sequence (SFMS) \citep{Popesso2023}.
In contrast, the less massive but more highly star-forming southern companion lies on the SFMS\null.
The southern companion has more dust attenuation than the northern one, despite the northern companion's higher mass and older stellar population. Higher dust content is consistent with more gas fuelling star formation.

\section{Foreground lensing source}
\label{sec:Lens}
\begin{figure*}
\begin{center}
\includegraphics[scale=0.86]{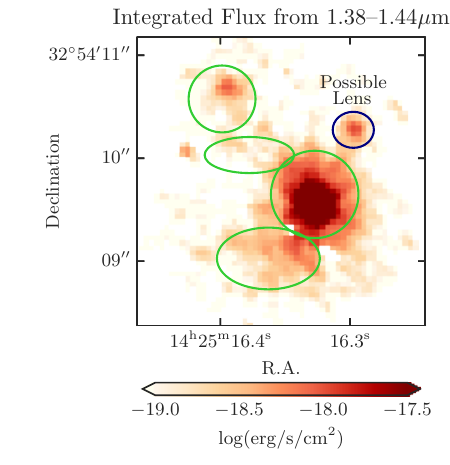}
\includegraphics[scale=0.86]{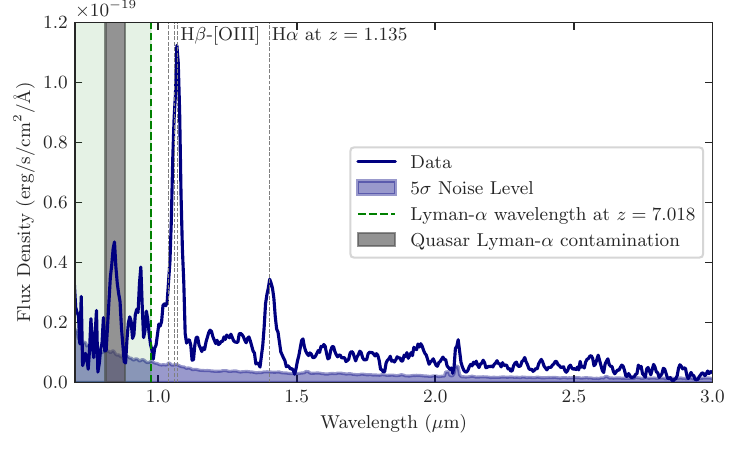} 
\caption{\textit{Left:} Flux from the full data cube integrated across 1.38--1.44\,\micron. The emission-line region apertures from Figure~\ref{fig:HST} are included for reference (green). A fifth aperture (black) marks the additional  source 0\farcs81 to the north-north-west of the quasar.
\textit{Right:}  Spectrum integrated over the black aperture in the left panel (blue). 
Three clear emission lines are seen at $0.85\,\micron$, 1.07\,\micron, and 1.40\,\micron. 
The $0.85\,\micron$ line is contaminating Lyman-alpha flux from the nearby quasar (grey shaded wavelengths).
There are two potential redshift solutions for the 1.07\,\micron\ and 1.40\,\micron\ emission lines: $z=1.135$ for the \oiiia\ and \ha\ emission lines (dashed grey lines) or $z=7.018$ from [\ions{C}{ii}]\,$\lambda$1334 and \ions{N}{iii}]$\,\lambda$1747. The green dashed line marks the Lyman-$\alpha$ redshift at $z=7.018$, below this in the green shaded region we would expect no flux for a high-$z$ galaxy. In fact, this region shows significant continuum flux, suggesting that the source is at $z=1.135$.
} 
\label{fig:Lens}
\end{center}
\end{figure*}

The IFU data cube shows an additional source 0\farcs81 north-north-west of the quasar, discovered via its two bright emission lines at 1.07\,\micron\ and 1.40\,\micron. 
Figure~\ref{fig:Lens} shows the location of this source and its extracted spectrum. There is clear continuum emission between ${\sim}$1 and 3\,\micron, with the emission also detectable in the HST/WFC3 F125W and F160W images with $m_J=28.8$\,mag and $m_H=26.9$\,mag.
There is an additional emission line present in the spectrum at $0.85\,\micron$, but this emission is purely contamination from the nearby quasar's Lyman-$\alpha$.
Given the spatial proximity to the quasar, this source could potentially be a foreground galaxy that may act as a gravitational lens for the quasar or, alternatively, this could be a background galaxy that could be lensed by the quasar. 

The most likely redshift solution for the two emission lines is $z=1.135\pm0.010$, which would identify the lines as \oiiia\ and \ha, respectively.
The asymmetry of the 1.07\,\micron\ line is consistent with being a blend between the doublet \oiii\ and potentially \hb, providing additional support for this redshift solution.
Unfortunately, these lines fall in the spectral region with the lowest resolution, $R\simeq35$ or $v_{\rm{inst}}\simeq8400\pm800$\,km\,s$^{-1}$. Combined with the issues that the 1.07\,\micron\ line would be a blend of lines and that \ha\ could be blended with \nii, the current data cannot measure the physical line widths of this system. Higher resolution spectra would give a velocity measurement of this galaxy as well as verifying the line identifications.

If the presumed foreground galaxy has significant mass and resides at $z=1.135$, it will gravitationally lens the background quasar.
If the lensing galaxy is a singular isothermal sphere with stellar velocity dispersion $\sigma$ between 50 and 145\,\kms, the quasar would always lie outside the caustic in the source plane and therefore not be multiply imaged. Such configurations have magnifications $1<\mu<2$ or $<0.75$\,mag.
If the galaxy's stellar velocity dispersion $\sigma \gtrsim 150$\,km\,s$^{-1}$, the lens would produce a secondary image of the quasar. However, we do not see a secondary image in the deep observations available, and thus such high velocity dispersion is ruled out.
Because the lens magnification of the quasar and the companions should be small, we have not applied any lensing correction. Precise line-width measurements for the lens could determine the actual amount of lensing.

The spectrum admits an alternative redshift solution $z=7.018\pm0.010$ for the north-west galaxy. In this case, the 1.07\,\micron\ line would come from [\ions{C}{ii}]$\,\lambda$1334 and the 1.40\,\micron\ line from \ions{N}{iii}]$\,\lambda$1747.
This scenario is unlikely because we would also expect to see the typically much brighter galaxy lines such as \ions{Mg}{ii}$\,\lambda$2799, [\ions{O}{ii}]$\,\lambda$3727, \oiii, and the hydrogen Balmer series, yet no additional emission lines are detected; the presence of only two emission lines and the blue continuum are more consistent with the $z=1.135$ solution.
Furthermore, if this source was at $z=7.018$, we would expect to see no continuum flux at wavelengths below the Lyman-$\alpha$ break at 0.975\,\micron. The noise in the IFU spectrum significantly increases at the bluest wavelengths, making this measurement difficult. However, summing the flux across the wavelength region 0.7--0.975\,\micron, excluding wavelengths 0.81--0.88\,\micron\ which contain flux from the quasar's Lyman-$\alpha$ line, gives a positive detection $f_{<0.975\micron}=(5.3\pm0.6)\times10^{-17}$ erg\,s$^{-1}$\,cm$^{-2}$.
The presence of statistically significant flux blue-wards of Lyman-$\alpha$ makes it unlikely that the source is at $z=7.018$, favouring the $z=1.135$ solution.

If this source is a background galaxy at $z=7.018$, it would likely be gravitationally lensed by the foreground quasar. 
We estimate the quasar host's stellar velocity dispersion from our black hole mass measurement using the \citet{Greene2020} $M_{\rm{BH}}$--$\sigma_\ast$ scaling relation.
Assuming that the quasar host system is a singular isothermal sphere with $\sigma_\ast=310\substack{+140\\-100}$\,km\,s$^{-1}$, we calculate that the background galaxy would not be multiply imaged, with magnifications of $1<\mu<2$, or $<0.75$\,mag. However, we reiterate that the $z=1.135$ solution is more likely, in which case it is this galaxy that is weakly lensing the background quasar.
A higher-resolution spectrum is needed to definitively confirm the redshift of this galaxy.

\section{Discussion}
\label{sec:discussion}
\subsection{Host stellar mass}
In this work, we have been unable to estimate the stellar mass of the
host galaxy, labelled as region~3. As discussed in Section~\ref{sec:SPSmodelling}, we could not perform SPS modelling for the host, as this requires a spectrum of the galaxy continuum emission across a broad wavelength range, which is not possible to produce with our quasar subtraction technique.
Developing a quasar-subtraction strategy that allows for stellar mass measurements will require significant additional work and will be the focus of a future study.

There are two existing mass estimates for the quasar host in the literature.
From the CO emission, \citet{Wang2010} estimated a dynamical mass $M_{\rm{dyn}}\geq1.56\times10^{11}$\,M$_\odot$.
However, the angular resolution of their PdBI observations was $\rm FWHM \sim5$\arcsec, making this estimate very uncertain.
An upper limit on the host mass came from non-detections of the host in the PSF-subtracted $J$- and $H$-band HST imaging \citep{Marshall2019c}. With significant assumptions, \citeauthor{Marshall2019c} placed an upper limit $M_{\ast}<1.5\times10^{11}$\,M$_\odot$. 
These measurements likely imply that the quasar host galaxy is less massive than the northern companion and
probably as massive as the southern companion (Table~\ref{tab:PhysicalProperties}). 
However, both observations and resulting mass measurements have significant limitations and these mass limits are difficult to interpret in the context of the companion galaxies.

The lack of accurate host stellar mass measurement results in two significant drawbacks to the current interpretation of \NDWFS. 
Firstly, we are limited in understanding the merger physics for this system. 
Without a host-galaxy mass measurement, it is unclear whether the mergers with the northern and southern companion are `major' mergers and whether the host is the most massive galaxy in this merging system. 
Statistically, we may expect that the quasar would reside in the most massive galaxy in the system \citep{Keel1996}, although the current stellar mass limit suggests that this is not the case.

Secondly, we cannot measure the black hole-stellar mass ratio for this quasar.
Observations with JWST \citep{Yue2023,Kokorev2023,Juodzbalis2024,Maiolino2023,Marshall2024} have been reporting high-$z$ black holes that lie significantly above the local relations and it would be interesting to see whether the black hole in this extreme merging system is similarly overmassive. 
The local black hole--stellar mass relation from \citet{Greene2020} says that a black hole with our measured $M_{\rm{BH}}=(1.4\substack{+3.1\\-1.0})\times10^9M_\odot$ will have a host galaxy with  $M_{\ast}=(3.6\substack{+10.4\\-0.9})\times10^{11}$\,\Msol, similar in mass to the northern companion.
The \citet{Marshall2019c} upper limit of $M_{\ast}<1.5\times10^{11}$\,M$_\odot$ suggests that \NDWFS\ will lie above the local relation, hosting an overmassive black hole.
An extreme black hole--stellar mass ratio similar to other high-$z$ quasars of ${\sim}10$--50\% \citep{Yue2023,Marshall2024} would result in $M_{\ast}\simeq0.3$--$1.4\times10^{10}\,M_\odot$, less massive than even the southern companion galaxy.
Now that we have an accurate Balmer-based black hole mass measurement, a stellar-mass measurement (or tighter upper limit) is needed to determine whether \NDWFS\ hosts a similarly overmassive black hole.
A better mass estimate will require better subtraction of the quasar flux or some indirect method.

\subsection{Merger dynamics}

The lack of an accurate stellar mass for the quasar host means we cannot determine which galaxy is the most massive `central' galaxy in the \NDWFS\ system. 
The \citet{Marshall2019c} upper limit for the host ($M_{\ast}<1.5\times10^{11}$\,\Msol) makes it most likely that the northern companion galaxy 
with $M_{\ast}=(3.6\substack{+0.6\\-0.3})\times10^{11}\,M_\odot$ 
is the central galaxy. This section describes the merger dynamics assuming that this is the case. However, the quoted dynamics are all relative, and 
the same analysis applies.

The velocity offset of $-820\pm217$\,km/s between the quasar host and the northern companion (Table~\ref{tab:HostFlux}) can have a contribution from two components: Hubble flow with the companion at a different cosmological distance than the host (i.e. a different cosmological redshift) and a physical (`peculiar') velocity of the companion relative to the quasar.
If there is no peculiar velocity, the northern companion would be 1.2~pMpc nearer than the quasar in the adopted cosmology.
However, the clear presence of a connecting gas bridge between the quasar and companion rules out this scenario. The velocity of the gas bridge varies smoothly between the quasar and companion, connecting the two, which would require the gas bridge to be over 1~Mpc long---clearly unphysical.
Indeed, requiring a physical separation $d_z<100$\,kpc, a generous maximum for a connecting-bridge scenario, requires a cosmological redshift difference no more than 1/12 of the actual redshift difference.
Therefore, nearly all of the ${\sim}820$\,km/s velocity offset between the companion and quasar must be a physical velocity, with the two at a similar cosmological distance.

The \citet{Cool2006} discovery spectrum has a significant decrease in flux from ${\sim}8360$--8380\,\AA, between the Lyman-$\alpha$ and \ions{N}{v}\,$\lambda1240$ line. \citet{Mechtley2014} determined that this is most likely foreground \hbox{\ions{H}{i}}\,$\lambda$1216\,\AA\ absorption from a companion galaxy, although alternate explanations including blueshifted \ions{N}{v} are possible.
Such an absorption wavelength corresponds to $z=5.875$--5.891 or between $-1100$ and $-400$\,\kms\ relative to the quasar host at $z=5.901$,
consistent with the measured velocity of the northern companion. Therefore the absorption feature is likely from the northern companion galaxy which lies in the foreground, with the quasar moving away from both us and the companion at ${\sim}820$\,\kms.
This suggests that the quasar was once in front of the companion galaxy, fell in towards the companion, and has now made its first pass and has begun to travel away.

The dynamical timescale at $z=5.9$ is ${\sim}150$\,Myr, while the age of the Universe at $z=5.9$ is 965~Myr. 
The expected merger timescale depends strongly on mass ratio and orbital eccentricity but will typically be of the order of the dynamical time \citep{CJiang2008,Boylan2008}, as will the dynamical-friction timescale \citep{Binney1987}.
The stellar ages (Table~\ref{tab:PhysicalProperties}) for both companions 
are significantly less than the dynamical time. This implies that the star formation in the galaxies 
most likely began while the galaxy interaction was in progress.
Lower-$z$ studies have found that SFRs are significantly enhanced at pair separation distances of $\lesssim$25\,kpc \citep{Shah2022} or potentially hundreds of kpc \citep{Patton2020}.
With a current physical separation $d_z<100$\,kpc, and with the quasar likely travelling only $\sim$55\,kpc in the $\sim65$\,Myr since the northern companion's star formation began (assuming constant speed of $820$\,\kms), the pair separation at the onset of the star formation was within this region of potential interaction-induced star formation.
Therefore, star formation in the companion having been triggered by the merger is a plausible scenario.
Indeed, it is also possible that the interaction has induced the quasar activity \citep{ByrneMamahit2024}.

The presence of the gas bridge provides further evidence that the quasar has already been through its first pass by the northern companion. 
As the quasar approached the companion, the interaction triggered an episode of star formation in the companion. After the close fly-by, the gas was tidally torn away from the galaxy outskirts, creating a tidal bridge \citep{Toomre1972,Wright1972,Oh2008}. The bridge is clear evidence of an ongoing interaction between the northern companion and the quasar host galaxy.

The relative dynamics of the southern companion are more difficult to constrain, given the lack of a connecting gas bridge. However, this companion has a negligible velocity offset from the quasar host of $36\pm217$\,km/s, strongly suggesting that it is associated with the quasar.

These various interacting galaxies must live the same dark-matter halo.
One way to estimate the halo mass is from the stellar mass of the central galaxy, either the northern companion or the host if it is more massive. Applying the stellar--halo mass relation \citep{Girelli2020,Shuntov2022} to the northern companion stellar mass gives a halo-mass estimate $M_h\simeq10^{12.5}$--$10^{13}$\,M$_\odot$. This implies a virial velocity of 500--650\,km/s, and so the low-velocity southern neighbour would be strongly bound to the quasar. 
To bind the northern neighbour and quasar host, the virial velocity must be ${\ga}820\pm217$\,\kms, consistent with a massive halo with $M_h\ga10^{13}$\,\Msol.
The inner core of the halo contains three merging galaxies, one of which hosts a luminous quasar. This is likely a protocluster core that will evolve into an extremely massive $\gtrsim10^{15}$\,\Msol\ cluster by $z=0$ \citep[e.g.][]{Reed2003}.

\subsection{High-z quasars in galaxy mergers}
\NDWFS\ joins a growing list of high-$z$ quasars that are spectroscopically confirmed to be undergoing galaxy--galaxy mergers.
PSO J308.0416$-$21.2339
at $z\simeq6.2$ \citep{Decarli2019} is accreting two massive $M_\ast\simeq10^{10}$\,\Msol\ companion galaxies onto a quasar host with dynamical mass $M_{\rm{dyn}}>10^{11}$\,\Msol. One of the companion galaxies is connected to the quasar via a bridge clearly detected in \ha, \hb, and \oiii\ \citep{Loiacono2024,Decarli2024}.
DELS J0411$-$0907 at $z=6.82$ has a companion galaxy with $M_{\rm{dyn}}\simeq2.5\times10^{10}$\,\Msol\ merging with an $M_{\rm{dyn}}\simeq8.6\times10^{10}$\,\Msol\ quasar host \citep{Marshall2023}.
VDES J0020$-$3653 at $z=6.86$ is merging with three companions, two of which have $M_{\rm{dyn}}\simeq1.5\times10^{11}$\,\Msol\ and $\simeq8\times10^{10}$\,\Msol\ compared to host $M_{\rm{dyn}}\simeq1.7\times10^{11}$\,\Msol\ \citep{Marshall2023}.
ULAS J112001.48+064124.3
at $z=7.08$ is also merging with a bright companion galaxy, with $M_{\ast}\simeq2.6\times10^9M_\odot$ for the host and $M_{*}\simeq5.0\times10^9M_\odot$ for the companion galaxy \citep{Marshall2024}.
BR~1202-0725 at $z\simeq4.7$ is in a highly overdense system, with an abundance of nearby galaxies on kpc scales including a secondary AGN \citep{Zamora2024}.
The quasar 
SDSS J113308.78+160355.7
at $z=5.63$ has a prominent bridge between the host and a dust-obscured companion as revealed with ALMA \citep{Zhu2024}. In this case, the companion is undergoing intense star formation of ${\sim}10^3$\,\Msol\,yr$^{-1}$.
\citet{Neeleman2019} studied five quasars with ALMA that have companions at separations of 5--60 kpc;
PSO J167.6415$-$13.4960
at $z=6.51$ and SDSS J130608.25+035626.3 at $z=6.03$ have connecting gas bridges between the companion and the quasar host.
At $z=6.05$ there is a pair of two merging quasars, separated by (projected) 12~kpc and with a connecting gas bridge \citep{Matsuoka2024,Izumi2024}.
The extremely luminous galaxy 
WISE J224607.55$-$052634.9,
which hosts a dust-obscured quasar at $z=4.6$, is connected to multiple merging companion galaxies via dust streams \citep{DiazSantos2018}.
The presence of gas bridges in all of these systems provides unambiguous evidence of significant ongoing interactions at high-$z$.

The great number of detections of merging high-$z$ quasars suggests that the intense black hole growth may be induced or enhanced by the merger activity.
Theoretically, mergers are expected to enhance both AGN and star formation 
\citep[e.g.][]{Doyon1994,Hernquist1995,Mihos1996,Hopkins2006,Patton2020,ByrneMamahit2023,ByrneMamahit2024}.
However, it is possible that the merger and the extreme black hole growth are unrelated and purely coincidental.
Similar extreme environments, with gas bridges extending between star-forming and dusty galaxies, are seen at similar redshifts for non-quasar galaxies \citep[e.g.][although these may show a different phase of the merger before or after induced quasar activity]{Litke2019,Asada2023,Villanueva2024,Solimano2025,Hu2025}.
A detailed investigation of the fraction of both quasars and similar galaxies undergoing mergers at these redshifts is needed for a clearer conclusion on the link between mergers and early black hole growth.

\subsection{Quasar--companion cross-ionisation}
\label{sec:crossion}

It has been observed in other systems that a central AGN may photoionise nearby companion galaxies \citep[so called cross-ionisation; e.g.][]{Moran1992,daSilva2011,Merluzzi2018,Keel2019,Moiseev2023,Protusova2024}.
The intrinsic \oiiia\ surface brightness in the \NDWFS\ extended regions is a close match on ${\sim}10$\,kpc scales to that in the extended emission-line regions in Mrk~266, which has the highest surface brightness extended line emission seen in the local Universe \citep{Keel2012b}.
This similarity raises the question of whether the quasar could be the source of ionisation for these nearby regions.
Each of the companion regions is indeed located in the region of the BPT diagnostic (Section~\ref{sec:BPT})
that could be explained by low-metallicity/high-$z$ AGN photoionisation \citep[e.g.][]{Groves2006,Nakajima2022,Dors2024,Cleri2025}. Therefore, we need to consider whether the central quasar has sufficient luminosity to produce this photoionisation.

For each of the companions, the ionising photon rate required to produce the observed \hb\ luminosity is 
\begin{equation}
Q_{\rm{comp}}=L_{\rm{H}\beta}/(4.8\times10^{-13}\rm{\,erg}),\quad
\end{equation}
assuming that one \hb\ photon (with energy $4.09\times10^{-12}$\,erg) is produced for every 8.54 recombinations \citep{Hummer1987}.
This gives required incident ionising-photon rates of 
${<}1.2\times10^{54}$\,s$^{-1}$ (3$\sigma)$,  $3.1\times10^{54}$\,s$^{-1}$, and $8.1\times10^{54}$\,s$^{-1}$
for regions 1, 2, and 4, respectively.

The incident ionising luminosity from the quasar onto a companion can be estimated as
\begin{equation}
L_{\rm{q, incident}} = L_{\rm{bol}} \gamma \frac{\Omega}{4\pi},\quad
\end{equation}
where $\gamma$ is the ratio of ionising to bolometric luminosity, and $\Omega$ is the solid angle the companion subtends as seen from the quasar. For small angles, $\Omega=\pi(\arctan{(r/d)})^2$, assuming a spherical galaxy of radius $r$ at distance $d$. For a typical quasar SED, $\gamma=0.14$ \citep{Keel2019}.
The extent of the \oiiia\ flux (Figure~\ref{fig:HST}) perpendicular to the direction of the quasar gives $r=0\farcs21$, 0\farcs25, and 0\farcs36 for regions 1, 2, and 4, respectively. 
We use the distance $d$ from the centre of each companion to the centre of the quasar measured from S\'ersic profile fits to the HST imaging \citep{Marshall2019c}, $d=1\farcs4$ and 0\farcs6 for regions 1 and 4, respectively. The connecting bridge extends from the quasar to region~1, with an average distance of $d=0\farcs7$. 
To convert ionising luminosity to an incident photon rate, we assume a mean ionising photon energy of 2~rydbergs or $4.36\times10^{-11}$\,erg, such that
\begin{equation}
Q_{\rm{q, incident}} = L_{\rm{q, incident}}/(4.36\times10^{-11}\,\rm{erg}).\quad
\end{equation}
For the three regions, this gives $Q_{\rm{q, incident}}=1.1\times10^{54}$\,s$^{-1}$, $5.3\times10^{54}$\,s$^{-1}$, and $13.6\times10^{54}$\,s$^{-1}$, respectively.

The incident photon rates from the quasar are at first glance sufficient to photoionise the two companions and the connecting bridge, with ratios of $Q_{\rm{q, incident}}/Q_{\rm{comp}} >0.9$ for region 1 and $Q_{\rm{q, incident}}/Q_{\rm{comp}}=1.7$ for both regions 2 and 4.
However, these calculations assume the minimum possible distances based on the sky projection. De-projected distances will likely require higher ionisation luminosities. These calculations also do not include dust attenuation corrections.
However, even if the true $Q_{\rm{q, incident}}/Q_{\rm{comp}}<1$, 
the quasar could contribute ionisation in addition to that produced by star formation.
Another possibility is that the quasar was previously more luminous and has faded, as seen for Hanny's Voorwerp and other AGN-ionised clouds on timescales of ${\sim}10^4$--$10^5$\,yr. \citep{Lintott2009,Schawinski2010,Keel2012b,Keel2012,Schirmer2013}. 

With the available diagnostics, it is plausible that the companion galaxies are photoionised by either a low-metallicity AGN, by star formation, or by a combination of the two. The quasar luminosity appears sufficient to ionise the surrounding regions out to kpc scales and the companions' sizes and emission-line surface brightnesses are comparable to local cross-ionised sources. The cross-ionisation of the companions from the quasar is therefore plausible (and perhaps likely). However,
additional emission-line diagnostics \citep[e.g.][]{Mignoli2013,Dors2024} are required for a definitive conclusion to be drawn.

\section{Conclusions}
\label{sec:Conclusions}

This work presents JWST NIRSpec IFU observations of the $z=5.89$ quasar \NDWFS, taken as part of the PEARLS GTO programme. The prism spectra cover the wavelength range  0.6--5.3\,\micron\ with a spectral resolution $R\sim100$ across the $3\arcsec\times3\arcsec$ FoV\null.
We estimated the NIRSpec IFU PSF and LSF (Appendix~\ref{sec:PSF} and~\ref{sec:LSF}) to accurately study the quasar system.

Scaling relations based on the \ha\ and \hb\ broad emission lines give a black hole mass of $M_{\rm{BH}}=(1.4\substack{+3.1\\-1.0})\times10^9\,M_\odot$, implying a high accretion rate over its history. However, \NDWFS\ is currently accreting below the Eddington limit, with $L_{\rm{Bol}}/L_{\rm{Edd}}=0.3\substack{+0.6\\-0.2}$. 
The quasar's spectrum around \oiii\ implies that there may be a quasar-driven outflow with velocity 
$6050\substack{+460\\-630}$\,km\,s$^{-1}$ and an ionised outflow rate of $1650\substack{+130\\-1230}$\,\Msol\,yr$^{-1}$, derived from \hb. This may be one of the most extreme outflows in the early Universe.
The stellar mass of the quasar host galaxy cannot be determined because of the quasar's light; however, based on the black hole--stellar mass relation, the host is likely to have $M_\ast\simeq10^{10}$--$10^{11}\,M_\odot$.
The quasar host likely has a significant amount of obscured star formation, with SFR$_{\rm{FIR}}=1160\,M_\odot$\,yr$^{-1}$ \citep{Wang2010} and SFR$_{\rm{H}\alpha}\simeq50$--$60\,M_\odot$\,yr$^{-1}$.

The IFU observations reveal that \NDWFS\ is part of a complex, merging system.
Other members of the system include two companion galaxies with projected separations of 5--8~kpc from the quasar host.
The companion in the north has an older, very massive stellar population that is relatively metal-rich, with moderate ongoing star formation significantly below the star-forming main sequence. This companion has begun merging with the quasar host and a clear bridge of hot, ionised gas connects the two galaxies. 
A second, distinct merger is ongoing from the south with a companion galaxy that is younger and less massive than the northern one. The southern companion likely has greater ongoing star formation than the northern companion, lying on the star-forming main sequence, but remains metal-poor.
Both companions and the gas bridge are likely at least partially photoionised by the central quasar. 
With two separate galaxies merging into the quasar host, this is a `train-wreck' merger in progress.
These ongoing mergers may have triggered both the ongoing star formation and intense quasar activity.

Finally, there is a potential foreground galaxy at $z=1.135$, located 0\farcs81 from the quasar. 
This galaxy is expected to act as a gravitational lens to the background quasar, giving a magnification of $1<\mu<2$, $<0.75$\,mag.
The discovery of this source was enabled by the $R\sim100$ prism spectra, sacrificing the spectral resolution to cover the full NIRSpec wavelength range of 0.6--5.3\,\micron\ in a single exposure. 
High-$z$ quasar studies that use the $R\sim2700$ grating provide more accurate kinematic measurements, more ideal for studying the \hb--\ha\ emission lines \citep[e.g.][]{Marshall2023,Marshall2024, Liu2024, Decarli2024,Loiacono2024}.
However, the typical G395M/F290LP mode produces a spectrum only from 2.9 to 5.3\,\micron, and that would have missed the foreground source's emission lines at 1--1.5\,\micron. 
The prism spectra also allowed for detections of the continuum emission from
the companion galaxies across the rest-frame UV and optical,
which would have been too faint to detect in $R\sim2700$ spectra with
equivalent exposure time; thus enabling SED fitting and stellar
mass estimates.
Taking $R\sim2700$ spectra across the full 0.6--5.3\,\micron\ wavelength range, deep enough to detect the continuum emission, would require significantly more observing time.

This superb data has revealed a potential extreme quasar-driven outflow, multiple companions, and a gas bridge, as well as a likely foreground galaxy lens.
Nevertheless, higher spectral-resolution observations 
would be ideal for 
studying in greater detail the kinematics of this extreme `train-wreck' merging-quasar system.

\begin{acknowledgements}
We dedicate this paper to the memory of our colleague Glenn Schneider, a champion of high-contrast imaging across the Universe from solar eclipses to high-$z$ quasars.\\

We are extremely grateful to our Program NIRSpec Reviewer Alaina Henry and Program Coordinator Tony Roman, who were critical in ensuring that these observations were successfully executed. 
We thank Glenn Wahlgren and the JWST Help Desk team for their help understanding the NIRSpec LSF\null.
We thank Alex Cameron for advice on using \textsc{pPXF}.
We thank the anonymous referee for a thoughtful review that helped to improve this paper.\\

This work is based on observations made with the NASA/ESA/CSA James Webb Space
Telescope. The data were obtained from the Mikulski Archive for Space
Telescopes at the Space Telescope Science Institute, which is operated by the
Association of Universities for Research in Astronomy, Inc., under NASA
contract NAS 5-03127 for JWST. These observations are associated with JWST
programme 1176 as part of the Prime Extragalactic Areas for Reionization and Lensing Science (PEARLS) GTO programme, and programme 1492.
The specific observations analysed can be accessed via the MAST archive at \url{http://dx.doi.org/10.17909/hbnz-7j44}.\\

MAM acknowledges support by the Laboratory Directed Research and Development programme of Los Alamos National Laboratory under project number 20240752PRD1.
RAW, SHC, and RAJ acknowledge support from NASA JWST Interdisciplinary
Scientist grants NAG5-12460, NNX14AN10G and 80NSSC18K0200 from GSFC. Work by
CJC and NJA acknowledge support from the European Research Council (ERC) Advanced
Investigator Grant EPOCHS (788113). 
CNAW acknowledges funding from the JWST/NIRCam contract NASS-0215 to the
University of Arizona.
CC is supported by the National Natural Science Foundation of China, No. 11803044, 11933003, 12173045, the Chinese Academy of Sciences (CAS), through a grant to the CAS South America Center for Astronomy (CASSACA), and the China Manned Space Project with grant NO. CMS-CSST-2021-A05.
AZ acknowledges support by the Ministry of Science \& Technology, Israel, and by Grant No. 2020750 from the United States-Israel Binational Science Foundation (BSF) and Grant No. 2109066 from the United States National Science Foundation (NSF).
GF and JSBW were supported by the Australian Research Council Centre of Excellence for All Sky Astrophysics in 3 Dimensions (ASTRO 3D), through project CE170100013
JL acknowledges support from the Research Grants Council (RGC) of Hong Kong through  RGC/GRF 17302023.
MP acknowledges support through the grants PID2021-127718NB-I00 and RYC2023-044853-I, funded by the Spain Ministry of Science and Innovation/State Agency of Research MCIN/AEI/10.13039/501100011033 and El Fondo Social Europeo Plus FSE+.
H\"U acknowledges funding by the European Union (ERC APEX, 101164796). Views and opinions expressed are however those of the authors only and do not necessarily reflect those of the European Union or the European Research Council Executive Agency. Neither the European Union nor the granting authority can be held responsible for them.\\

This research has made use of NASA's Astrophysics Data System, developed by Smithsonian Astrophysical Observatory.

This paper made use of Python packages and software
\textsc{AstroPy} \citep{Astropy2013},
\textsc{jwst} \citep{Bushouse2022},
\textsc{Matplotlib} \citep{Matplotlib2007},
\textsc{NumPy} \citep{Numpy2011},
\textsc{Pandas} \citep{reback2020pandas}, 
\textsc{Photutils} \citep{photutils},
\textsc{Regions} \citep{Bradley2022},
\textsc{SciPy} \citep{2020SciPy-NMeth},
\textsc{Seaborn} \citep{Waskom2021},
\textsc{Spectral Cube} \citep{Ginsburg2019}, 
\textsc{QDeblend3D} \citep{Husemann2013,Husemann2014}, 
\textsc{QubeSpec}\footnote{\url{https://github.com/honzascholtz/Qubespec}},
and
\textsc{WebbPSF} \citep{Perrin2015},
as well as the software QFitsView \citep{Ott2012}.
\end{acknowledgements}

\bibliography{aa54307-25}
\bibliographystyle{aa}

\begin{appendix}

\section{Point spread function}
\label{sec:PSF}
\begin{figure}[h!]
\begin{center}
\includegraphics[scale=0.86]{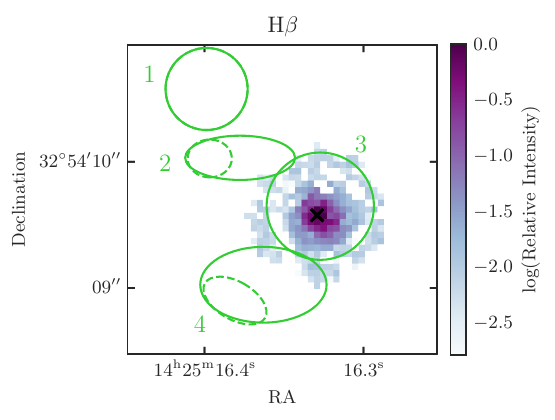}
\includegraphics[scale=0.86]{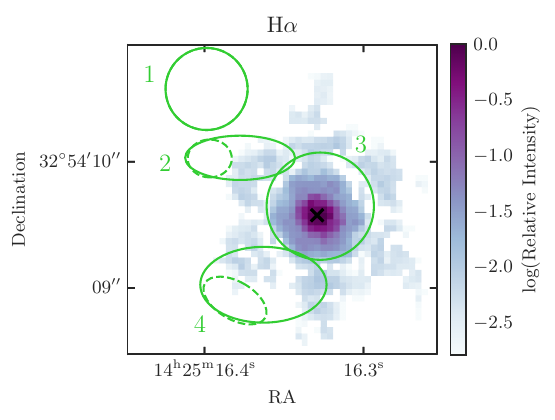}
\caption{NIRSpec IFU instrumental PSF at 3.35\,\micron\ as measured from the broad \hb\ line (top), and at 4.52\,\micron\ as measured from the broad \ha\ line (bottom) of \NDWFS. The coloured solid ellipses mark the emission line regions as in Figure \ref{fig:HST} to assist in visualising the PSF scale.
As regions 2 and 4 overlap the quasar PSF, we define two alternate apertures, regions 2* and 4*, within the original apertures but avoiding the region with quasar flux, shown as dashed ellipses.
This allows us to extract and model the continuum emission from these regions (Sections~\ref{sec:SPSmodelling} and~\ref{sec:pPXF}) without quasar contamination.}
\label{fig:PSFs}
\end{center}
\end{figure}

Using \textrm{QDeblend3D} \citep{Husemann2013,Husemann2014}, we determine the PSF shape by measuring the flux of the quasar BLR wings in each spaxel. 
As the PSF varies with wavelength, we measure these BLR fluxes for both the \hb\ and \ha\ broad BLR lines.
For \ha, we measure the flux across rest-frame 6485--6525\,\AA\ and 6605--6640\,\AA, i.e., both sides of the emission peak at 6564.6\,\AA.
For \hb, we measure the flux across rest-frame 4800--4850\,\AA, i.e., only on the blue side of the peak at 4862.68\,\AA, as the red side is contaminated by blending with \oiiib.
Figure~\ref{fig:PSFs} shows the resulting 2D \hb\ and \ha\ BLR flux maps, that is, the IFU PSF shape at 3.35 and 4.52\,\micron.
The FWHM of each PSF, averaged in the horizontal and vertical directions, is 0\farcs14 at 3.35\,\micron\ and 0\farcs18 at 4.52\,\micron. These measurements are consistent with the PSF FWHM measurements for the NIRSpec IFU from \citet{D'Eugenio2024}.

\section{Line spread function}
\label{sec:LSF}

\begin{figure*}
\begin{center}
\includegraphics[scale=0.86]{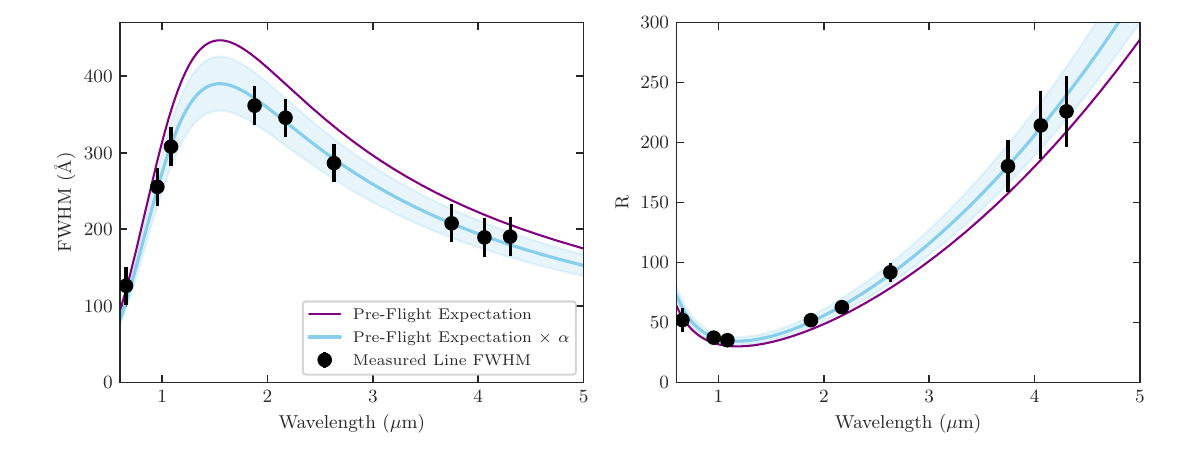}
\caption{NIRSpec IFU prism instrumental FWHM (left) as a function of wavelength. The purple curve shows the pre-flight estimates. The black points are measurements of emission lines from the planetary nebula SMP LMC 58, which have negligible intrinsic line width. The blue curve is the optimal scaling of the pre-flight expectation to match the observed measurements of the planetary nebula, with the shaded region showing the $\pm1\sigma$ uncertainty. The right panel shows the corresponding spectral resolving power $R=\lambda/\rm{FWHM}$.}
\label{fig:LSF}
\end{center}
\end{figure*}

Throughout this analysis, we measured line FWHMs that were smaller than expected from the pre-flight estimate of the NIRSpec prism LSF ($R\sim35$--300). Accurately measuring line widths and velocities requires a more accurate estimate of the in-flight LSF.

The compact planetary nebula SMP LMC 58 = IRAS 05248$-$7007 was observed in the JWST commissioning and calibration programmes PID 1125 and 1492 to characterise the NIRSpec LSF \citep[see e.g.][who performed a similar analysis on SMP LMC 58 NIRSpec data from other filter--grating pairs]{Isobe2023}. 
We download the IFU prism data from the planetary nebula from PID 1492, as reduced by the default MAST pipeline with calibration software version 1.14.0 and reference file \textsc{jwst\_1253.pmap}. We use the prism data from Observation~8, which has 4 dithers in a 4-point nod pattern and a total exposure time of 350\,s. We integrated the spectrum across a circular aperture of radius 1\arcsec\ surrounding the centre of the emission, and subtracted the continuum to obtain a pure emission-line spectrum.
We measure the FWHM of the planetary nebula's emission lines by fitting each with a Gaussian profile. We ignore lines with low $S/N$ or blended with other lines, leaving nine emission lines from 0.6--4.3\,\micron. 
The emission lines of this planetary nebula were unresolved in VLT/X-shooter $R\sim6500$ spectra from 1\farcs2 slits \citep{EuclidPN2023,Isobe2023} and therefore will also be unresolved in this $R\sim100$ spectra. We therefore assume that the observed line width arises purely from instrumental broadening. Figure~\ref{fig:LSF} shows the measured instrumental FWHM as a function of wavelength. 
Uncertainties in $\rm{FWHM}(\lambda)$ are calculated by assuming that each line's FWHM is measured with an uncertainty $\sigma$ of half a wavelength element, $\sigma=25$\,\AA.

Owing to the limited number of emission lines, we cannot estimate the shape of $\rm{FWHM}(\lambda)$. Instead, we use the shape of the spectral resolving power $R(\lambda)$ as estimated from the pre-flight data\footnote{as tabulated at \url{https://jwst-docs.stsci.edu/jwst-near-infrared-spectrograph/nirspec-instrumentation/nirspec-dispersers-and-filters}}, where $1/R\equiv\rm{FWHM}/\lambda$.
The optimal least squares solution for the constant $\alpha$ in the linear equation $\rm{FWHM}(\lambda)=\alpha~ \rm{FWHM}_{\rm{pre-flight}}(\lambda)$ is $0.873\pm0.079$.
The uncertainties are calculated by assuming that the FWHM measurements are all $\sigma=25$\,\AA\ larger or smaller than our best estimates.
Modifying the integration aperture by $\pm50\%$ adds an additional uncertainty of $\pm0.005$.

Translating these to measurements of the spectral resolving power gives $R(\lambda)=\beta~ R_{\rm{pre-flight}}(\lambda)$, where $\beta=1/\alpha=1.14\substack{+0.11\\-0.09}$. That is, we measure a spectral resolution that is $14\substack{+11\\-9}$\% larger than the pre-flight expectations.
Figure~\ref{fig:LSF} shows the resulting spectral resolving power as a function of wavelength. 
This factor of increase is broadly consistent with the assumed multiplication factors used in previous NIRSpec studies, for example, \citet{Greene2024} derived $\beta=1.3$, and \citet{Slob2024} derived $\beta=1.2$, although those were for the Micro-Shutter Assembly rather than the IFU\null.  \citet{Isobe2023}, \citet{deGraaff2024}, \citet{Nanayakkara2024}, and \citet{Shajib2025} have measured the NIRSpec resolution for other observing modes and filter--disperser pairs.

The corresponding velocity resolution is $\rm{FWHM}_{\rm{inst}}=c/R$, where $c$ is the speed of light.
At the \NDWFS\ redshift of $z=5.89$, the instrumental velocity resolutions for our key lines are 
$\rm{FWHM}_{\rm{inst}}=2083\pm188$\,km\,s$^{-1}$, $1997\pm181$\,km\,s$^{-1}$, $1966\pm178$\,km\,s$^{-1}$, and $1126\pm102$\,km\,s$^{-1}$, for \hb, \oiii, and \ha\ respectively. We assume that $\rm{FWHM}_{\rm{inst, [NII]}}=\rm{FWHM}_{\rm{inst, H}\alpha}$.
The stated uncertainties are based on the uncertainties in $\alpha$ above.

\end{appendix}

\label{LastPage}
\end{document}